%% dec. 4, 2002

\input harvmac
\input epsf
%\draftmode
\overfullrule=0pt

\def\IL{\relax{\rm I\kern-.18em L}}
\def\IH{\relax{\rm I\kern-.18em H}}
\def\IR{\relax{\rm I\kern-.18em R}}
\def\IC{\relax\hbox{$\inbar\kern-.3em{\rm C}$}}
\def\IZ{\relax\ifmmode\mathchoice
{\hbox{\cmss Z\kern-.4em Z}}{\hbox{\cmss Z\kern-.4em Z}}
{\lower.9pt\hbox{\cmsss Z\kern-.4em Z}}
{\lower1.2pt\hbox{\cmsss Z\kern-.4em Z}}\else{\cmss Z\kern-.4em
Z}\fi}

\def\CN {{\cal N}}
\def\CR {{\cal R}}
\def\CD {{\cal D}}

\def\CP {{\cal P }}
\def\CL {{\cal L}}
\def\CV {{\cal V}}
\def\CO {{\cal O}}

\def\CC {{\cal C}}
\def\CB {{\cal B}}
\def\CS {{\cal S}}

\def\CX{{\cal X}}

%% MORE MACROS

\def\CN {{\cal N}}

\def\CO {{\cal O}}

\def\CP {{\cal P }}

\def\CV{{\cal V }}

\def\CS {{\cal S }}

\font\manual=manfnt \def\dbend{\lower3.5pt\hbox{\manual\char127}}

\def\IZ{\relax\ifmmode\mathchoice
{\hbox{\cmss Z\kern-.4em Z}}{\hbox{\cmss Z\kern-.4em Z}}
{\lower.9pt\hbox{\cmsss Z\kern-.4em Z}}
{\lower1.2pt\hbox{\cmsss Z\kern-.4em Z}}\else{\cmss Z\kern-.4em
Z}\fi}
\def\half {{1\over 2}}

\def\p{\partial}

\def\CN {{\cal N}}

\def\CO {{\cal O}}

\def\CP {{\cal P }}

\def\CV{{\cal V }}

\def\CS {{\cal S }}

% more macros, alphabetically

\def\IZ{\relax\ifmmode\mathchoice
{\hbox{\cmss Z\kern-.4em Z}}{\hbox{\cmss Z\kern-.4em Z}}
{\lower.9pt\hbox{\cmsss Z\kern-.4em Z}}
{\lower1.2pt\hbox{\cmsss Z\kern-.4em Z}}\else{\cmss Z\kern-.4em
Z}\fi}
\def\IB{\relax{\rm I\kern-.18em B}}
\def\IC{{\relax\hbox{$\inbar\kern-.3em{\rm C}$}}}
\def\ID{\relax{\rm I\kern-.18em D}}
\def\IE{\relax{\rm I\kern-.18em E}}
\def\IF{\relax{\rm I\kern-.18em F}}
\def\IG{\relax\hbox{$\inbar\kern-.3em{\rm G}$}}
\def\IGa{\relax\hbox{${\rm I}\kern-.18em\Gamma$}}
\def\IH{\relax{\rm I\kern-.18em H}}
\def\II{\relax{\rm I\kern-.18em I}}
\def\IK{\relax{\rm I\kern-.18em K}}
\def\IP{\relax{\rm I\kern-.18em P}}
\def\IQ{\relax\hbox{$\inbar\kern-.3em{\rm Q}$}}

\def\lieg{{\underline{\bf g}}}

\def\inbar{\,\vrule height1.5ex width.4pt depth0pt}

\def\mod{{\rm mod}}
\def\p{\partial}

\font\cmss=cmss10 \font\cmsss=cmss10 at 7pt
\def\IR{\relax{\rm I\kern-.18em R}}

\def\vol{{\rm vol}}

% Macros for boxes
%
%\def\boxit#1{\vbox{\hrule\hbox{\vrule\kern8pt
%\vbox{\hbox{\kern8pt}\hbox{\vbox{#1}}\hbox{\k
%\hbox{$\displaystyle #1$}\kern8pt}\kern8pt\vrule}\hrule}}
%
%%% MACROS FOR BOX BOUNDARY CONDS
%%% FROM KAWAI ET AL

\def\makeblankbox#1#2{\hbox{\lower\dp0\vbox{\hidehrule{#1}{#2}%
   \kern -#1% overlap rules
   \hbox to \wd0{\hidevrule{#1}{#2}%
      \raise\ht0\vbox to #1{}% vrule height
      \lower\dp0\vtop to #1{}% vrule depth
      \hfil\hidevrule{#2}{#1}}%
   \kern-#1\hidehrule{#2}{#1}}}%
}%
\def\hidehrule#1#2{\kern-#1\hrule height#1 depth#2 \kern-#2}%
\def\hidevrule#1#2{\kern-#1{\dimen0=#1\advance\dimen0 by #2\vrule
    width\dimen0}\kern-#2}%
\def\openbox{\ht0=1.2mm \dp0=1.2mm \wd0=2.4mm  \raise 2.75pt
\makeblankbox {.25pt} {.25pt}  }

\def\bun#1/#2{\leavevmode
   \kern.1em \raise .5ex \hbox{\the\scriptfont0 #1}%
   \kern-.1em $/$%
   \kern-.15em \lower .25ex \hbox{\the\scriptfont0 #2}%
}

\def\opensquare{\ht0=3.4mm \dp0=3.4mm \wd0=6.8mm  \raise 2.7pt
\makeblankbox {.25pt} {.25pt}  }

%%%%%%%%%%%%%%%%%%%%%%%

\def\sector#1#2{\ {\scriptstyle #1}\hskip 1mm
\mathop{\opensquare}\limits_{\lower 1mm\hbox{$\scriptstyle#2$}}\hskip 1mm}

\def\tsector#1#2{\ {\scriptstyle #1}\hskip 1mm
\mathop{\opensquare}\limits_{\lower 1mm\hbox{$\scriptstyle#2$}}^\sim\hskip 1mm}
%%%
%%%

%% ANOTHER SET OF MACROS

\def\lieg{{\underline{\bf g}}}

\def\inbar{\,\vrule height1.5ex width.4pt depth0pt}

\def\p{\partial}

\font\cmss=cmss10 \font\cmsss=cmss10 at 7pt
\def\IR{\relax{\rm I\kern-.18em R}}

\def\vol{{\rm vol}}

%%%%%%%%%%%%%%%%%%%%%%%%%%%%%%%%%%%%%%%%%%%%%%%%%%%%%%%%%%%%%%%%%%
%%%%%%%%%%%%%%%%%%%%%%%%%%%%%%%%%%%%%%%%%%%%%%%%%%%%%%%%%%%%%%%%%%
%% new macros, specific to this paper
%%%%%%%%%%%%%%%%%%%%%%%%%%%%%%%%%%%%%%%%%%%%%%%%%%%%%%%%%%%%%%%%%%

\def\ie{{\it i.e.}}
\def\eg{{\it e.g.}}
\def\cf{{\it c.f.}}

\def\etc{{\it etc}}

\def\nextline{\hfil\break}

\def\sst{\scriptscriptstyle}

\def\frac#1#2{{#1\over#2}}
\def\coeff#1#2{{\textstyle{#1\over #2}}}
\def\half{\frac12}

\def\Xbar{\bar X}
\def\eff{{\rm eff}}
\def\kahler{K\"ahler}
\def\uu{{\bf u}}
\def\vv{{\bf v}}
\def\ww{{\bf w}}

\def\X{{\sst X}}
\def\Y{{\sst Y}}

%%%%%%%%%%%%%%%%%%%%%%%%%%%%%%%%%%%%%%%%%%%%%%%%%%%%%%%%%%%%%%%%%%%%%%%%%%
% figure names
%
\def\ring{1}
\def\glsm{2}
\def\rhobdy{3}
\def\genericflow{4}
\def\specialflow{5}
\def\rhoflow{6}
\def\newton{7}
%
%%%%%%%%%%%%%%%%%%%%%%%%%%%%%%%%%%%%%%%%%%%%%%%%%%%%%%%%%%%%%%%%%%%%%%%%%%

%% END MACROS
%%

%%%%%%%%%%%%%%%%%%%%%%%%%%%%%%%%%%%%%%%%%%%%%%%%%%%%%%%%%%%%%%%%%%%%%%%%%%
%%%%%%%%%%%%%%%%%%%%%%%%%%%%%%%%%%%%%%%%%%%%%%%%%%%%%%%%%%%%%%%%%%%%%%%%%%
%%% 
%%% References 
%%%
%%%%%%%%%%%%%%%%%%%%%%%%%%%%%%%%%%%%%%%%%%%%%%%%%%%%%%%%%%%%%%%%%%%%%%%%%%
%%%%%%%%%%%%%%%%%%%%%%%%%%%%%%%%%%%%%%%%%%%%%%%%%%%%%%%%%%%%%%%%%%%%%%%%%%

\lref\fulton{W. Fulton, {\it Introduction to Toric Varieties},
Annals of Mathematics Studies, vol. 131; Princeton Univ. Press (1993).} 
\lref\bpv{W.~Barth, C.~Peters, A.~Van de Ven, {\it Compact Complex Surfaces};
Springer-Verlag (1984).} 
\lref\stevens{J. Stevens, ``On the versal deformation
of cyclic quotient singularities'',
in {\it Singularity theory and its applications, part I},
LNM 1462 pp.302-319.}
\lref\ishii{A. Ishii, ``On McKay correspondence
for a finite small subgroup of GL(2,C)'',
to appear in J. Reine Ang. Math.
(available at \nextline
http://www.kusm.kyoto-u.ac.jp/preprint/preprint2000.html).}
\lref\wunram{J. Wunram, ``Reflexive modules on quotient surface
singularities'', Math. Ann. {\bf 279}, 583 (1988).}
\lref\riemenschneider{O. Riemenschneider,
``Special representations and the two-dimensional McKay correspondence''
(available at \nextline
http://www.math.uni-hamburg.de/home/riemenschneider/hokmckay.ps).}
%
%\AdamsSV
\lref\AdamsSV{
A.~Adams, J.~Polchinski and E.~Silverstein,
``Don't panic! Closed string tachyons in ALE space-times,''
JHEP {\bf 0110}, 029 (2001)
arXiv:hep-th/0108075.
%%CITATION = HEP-TH 0108075;%%
}
%
%\HarveyWM
\lref\HarveyWM{
J.~A.~Harvey, D.~Kutasov, E.~J.~Martinec and G.~Moore,
``Localized tachyons and RG flows,''
arXiv:hep-th/0111154.
%%CITATION = HEP-TH 0111154;%%
}
%
%\VafaRA
\lref\VafaRA{
C.~Vafa,
``Mirror symmetry and closed string tachyon condensation,''
arXiv:hep-th/0111051.
%%CITATION = HEP-TH 0111051;%%
}
%
%\MartinecTZ
\lref\MartinecTZ{
E.~J.~Martinec,
``Defects, decay, and dissipated states,''
arXiv:hep-th/0210231.
%%CITATION = HEP-TH 0210231;%%
}
%
%\MorrisonFR
\lref\MorrisonFR{
D.~R.~Morrison and M.~Ronen Plesser,
``Summing the instantons: Quantum cohomology 
and mirror symmetry in toric varieties,''
Nucl.\ Phys.\ B {\bf 440}, 279 (1995)
hep-th/9412236.
%%CITATION = HEP-TH 9412236;%%
}
%
%\CecottiVB
\lref\CecottiVB{
S.~Cecotti and C.~Vafa,
``Exact results for supersymmetric sigma models,''
Phys.\ Rev.\ Lett.\  {\bf 68}, 903 (1992)
hep-th/9111016.
%%CITATION = HEP-TH 9111016;%%
}
%
%\WittenYC
\lref\WittenYC{
E.~Witten,
``Phases of N = 2 theories in two dimensions,''
Nucl.\ Phys.\ B {\bf 403}, 159 (1993)
hep-th/9301042.
%%CITATION = HEP-TH 9301042;%%
}
%
%\HoriCK
\lref\HoriCK{
K.~Hori, A.~Iqbal and C.~Vafa,
``D-branes and mirror symmetry,''
arXiv:hep-th/0005247.
%%CITATION = HEP-TH 0005247;%%
}
%
%\HoriKT
\lref\HoriKT{
K.~Hori and C.~Vafa,
``Mirror symmetry,''
arXiv:hep-th/0002222.
%%CITATION = HEP-TH 0002222;%%
}
%
%\HoriFJ
\lref\HoriFJ{
K.~Hori,
``Mirror symmetry and some applications,''
arXiv:hep-th/0106043.
%%CITATION = HEP-TH 0106043;%%
}

%\HoriIC
\lref\HoriIC{
K.~Hori,
``Linear models of supersymmetric D-branes,''
arXiv:hep-th/0012179.
%%CITATION = HEP-TH 0012179;%%
}
%
%\HellermanBU
\lref\HellermanBU{
S.~Hellerman, S.~Kachru, A.~E.~Lawrence and J.~McGreevy,
``Linear sigma models for open strings,''
JHEP {\bf 0207}, 002 (2002)
arXiv:hep-th/0109069.
%%CITATION = HEP-TH 0109069;%%
}
%
%\LercheUY
\lref\LercheUY{
W.~Lerche, C.~Vafa and N.~P.~Warner,
``Chiral Rings In N=2 Superconformal Theories,''
Nucl.\ Phys.\ B {\bf 324}, 427 (1989).
%%CITATION = NUPHA,B324,427;%%
}
%
%\DouglasSW
\lref\DouglasSW{
M.~R.~Douglas and G.~W.~Moore,
``D-branes, Quivers, and ALE Instantons,''
arXiv:hep-th/9603167.
%%CITATION = HEP-TH 9603167;%%
}
%
%\HarveyNA
\lref\HarveyNA{
J.~A.~Harvey, D.~Kutasov and E.~J.~Martinec,
``On the relevance of tachyons,''
arXiv:hep-th/0003101.
%%CITATION = HEP-TH 0003101;%%
}
%
%\SenMD
\lref\SenMD{
A.~Sen,
``Supersymmetric world-volume action for non-BPS D-branes,''
JHEP {\bf 9910}, 008 (1999)
arXiv:hep-th/9909062.
%%CITATION = HEP-TH 9909062;%%
}
%
%\SenXM
\lref\SenXM{
A.~Sen,
``Universality of the tachyon potential,''
JHEP {\bf 9912}, 027 (1999)
arXiv:hep-th/9911116.
%%CITATION = HEP-TH 9911116;%%
}
%
%\KutasovQP
\lref\KutasovQP{
D.~Kutasov, M.~Marino and G.~W.~Moore,
``Some exact results on tachyon condensation in string field theory,''
JHEP {\bf 0010}, 045 (2000)
arXiv:hep-th/0009148.
%%CITATION = HEP-TH 0009148;%%
}
%
%\GerasimovZP
\lref\GerasimovZP{
A.~A.~Gerasimov and S.~L.~Shatashvili,
``On exact tachyon potential in open string field theory,''
JHEP {\bf 0010}, 034 (2000)
arXiv:hep-th/0009103.
%%CITATION = HEP-TH 0009103;%%
}

\lref\reid{
M.~Reid,
``La correspondance de McKay,''
S\'eminaire Bourbaki, 52\`eme ann\'ee, novembre 1999, no. 867, 
to appear in Ast\'erisque 2000
arXiv:alg-geom/9911165. For further references see 
http://www.maths.warwick.ac.uk/$\scriptstyle\sim$miles/McKay/
}

%\ReidZY
\lref\ReidZY{
M.~Reid,
``McKay correspondence,''
arXiv:alg-geom/9702016.
%%CITATION = ALG-GEOM 9702016;%%
}

%\MayrAS
\lref\MayrAS{
P.~Mayr,
``Phases of supersymmetric D-branes on Kaehler 
manifolds and the McKay  correspondence,''
JHEP {\bf 0101}, 018 (2001)
arXiv:hep-th/0010223.
%%CITATION = HEP-TH 0010223;%%
}

%\AnselmiSM
\lref\AnselmiSM{
D.~Anselmi, M.~Billo, P.~Fre, L.~Girardello and A.~Zaffaroni,
``Ale Manifolds And Conformal Field Theories,''
Int.\ J.\ Mod.\ Phys.\ A {\bf 9}, 3007 (1994)
arXiv:hep-th/9304135.
%%CITATION = HEP-TH 9304135;%%
}

%\BuscherQJ
\lref\BuscherQJ{
T.~H.~Buscher,
``Path Integral Derivation Of Quantum Duality In Nonlinear Sigma Models,''
Phys.\ Lett.\ B {\bf 201}, 466 (1988).
%%CITATION = PHLTA,B201,466;%%
}

%\RocekPS
\lref\RocekPS{
M.~Rocek and E.~Verlinde,
``Duality, quotients, and currents,''
Nucl.\ Phys.\ B {\bf 373}, 630 (1992)
arXiv:hep-th/9110053.
%%CITATION = HEP-TH 9110053;%%
}

%\DelaOssaXK
\lref\DelaOssaXK{
X.~De la Ossa, B.~Florea and H.~Skarke,
``D-branes on noncompact Calabi-Yau manifolds: K-theory and monodromy,''
Nucl.\ Phys.\ B {\bf 644}, 170 (2002)
arXiv:hep-th/0104254.
%%CITATION = HEP-TH 0104254;%%
}

\lref\ito{
Y.~Ito,
``Special McKay correspondence,''
arXiv:alg-geom/0111314.
}

\lref\morelli{
R. Morelli, ``K theory of a toric variety,'' Adv. in Math. {\bf 100}(1993)154
}

%\AspinwallXS
\lref\AspinwallXS{
P.~S.~Aspinwall and M.~R.~Plesser,
``D-branes, discrete torsion and the McKay correspondence,''
JHEP {\bf 0102}, 009 (2001)
arXiv:hep-th/0009042.
%%CITATION = HEP-TH 0009042;%%
}
%\DiaconescuEC
\lref\DiaconescuEC{
D.~E.~Diaconescu and M.~R.~Douglas,
``D-branes on stringy Calabi-Yau manifolds,''
arXiv:hep-th/0006224.
%%CITATION = HEP-TH 0006224;%%
}
%\DiaconescuBR
\lref\DiaconescuBR{
D.~E.~Diaconescu, M.~R.~Douglas and J.~Gomis,
``Fractional branes and wrapped branes,''
JHEP {\bf 9802}, 013 (1998)
arXiv:hep-th/9712230.
%%CITATION = HEP-TH 9712230;%%
}

%\LercheVJ
\lref\LercheVJ{
W.~Lerche, P.~Mayr and J.~Walcher,
``A new kind of McKay correspondence from non-Abelian gauge theories,''
arXiv:hep-th/0103114.
%%CITATION = HEP-TH 0103114;%%
}

%\WittenCD
\lref\WittenCD{
E.~Witten,
``D-branes and K-theory,''
JHEP {\bf 9812}, 019 (1998)
arXiv:hep-th/9810188.
%%CITATION = HEP-TH 9810188;%%
}
%\GarciaCompeanRG
\lref\GarciaCompeanRG{
H.~Garcia-Compean,
``D-branes in orbifold singularities and equivariant K-theory,''
Nucl.\ Phys.\ B {\bf 557}, 480 (1999)
arXiv:hep-th/9812226.
%%CITATION = HEP-TH 9812226;%%
}
%\DelaOssaXK
\lref\DelaOssaXK{
X.~De la Ossa, B.~Florea and H.~Skarke,
``D-branes on noncompact Calabi-Yau manifolds: K-theory and monodromy,''
Nucl.\ Phys.\ B {\bf 644}, 170 (2002)
arXiv:hep-th/0104254.
%%CITATION = HEP-TH 0104254;%%
}

%\GovindarajanVI
\lref\GovindarajanVI{
S.~Govindarajan and T.~Jayaraman,
``D-branes, exceptional sheaves and quivers on 
Calabi-Yau manifolds: From Mukai to McKay,''
Nucl.\ Phys.\ B {\bf 600}, 457 (2001)
arXiv:hep-th/0010196.
%%CITATION = HEP-TH 0010196;%%
}
%\GovindarajanEF
\lref\GovindarajanEF{
S.~Govindarajan, T.~Jayaraman and T.~Sarkar,
``On D-branes from gauged linear sigma models,''
Nucl.\ Phys.\ B {\bf 593}, 155 (2001)
arXiv:hep-th/0007075.
%%CITATION = HEP-TH 0007075;%%
}

%\HeCR
\lref\HeCR{
Y.~H.~He,
``On algebraic singularities, finite graphs and D-brane gauge theories:
A  string theoretic perspective,''
arXiv:hep-th/0209230.
%%CITATION = HEP-TH 0209230;%%
}

%\TakayanagiXT
\lref\TakayanagiXT{
T.~Takayanagi,
``Tachyon condensation on orbifolds and McKay correspondence,''
Phys.\ Lett.\ B {\bf 519}, 137 (2001)
arXiv:hep-th/0106142.
%%CITATION = HEP-TH 0106142;%%
}

%\TomasielloYM
\lref\TomasielloYM{
A.~Tomasiello,
``D-branes on Calabi-Yau manifolds and helices,''
JHEP {\bf 0102}, 008 (2001)
arXiv:hep-th/0010217.
%%CITATION = HEP-TH 0010217;%%
}

%\GovindarajanVI
\lref\GovindarajanVI{
S.~Govindarajan and T.~Jayaraman,
``D-branes, exceptional sheaves and quivers on Calabi-Yau manifolds:
{}From  Mukai to McKay,''
Nucl.\ Phys.\ B {\bf 600}, 457 (2001)
arXiv:hep-th/0010196.
%%CITATION = HEP-TH 0010196;%%
}

%\BatyrevJU
\lref\BatyrevJU{
V.~V.~Batyrev and D.~I.~Dais,
``Strong Mckay Correspondence, String Theoretic Hodge Numbers And
Mirror Symmetry,''
arXiv:alg-geom/9410001.
%%CITATION = ALG-GEOM 9410001;%%
}

%\ItoZX
\lref\ItoZX{
Y.~Ito and M.~Reid,
``The McKay correspondence for finite subgroups of SL(3,C),''
arXiv:alg-geom/9411010.
%%CITATION = ALG-GEOM 9411010;%%
}

\lref\itonak{
Y.~Ito and H.~Nakajima,
``McKay correspondence and Hilbert schemes in dimension three,''
arXiv:al-geom/9802120.
}

\lref\crawthesis{
A.~Craw,
``The McKay correspondence and representations of the McKay quiver,''
Ph.D. thesis, University of Warwick; available at
http://www.math.utah.edu/~craw.
}

%\KachruAN
\lref\KachruAN{
S.~Kachru, S.~Katz, A.~E.~Lawrence and J.~McGreevy,
``Mirror symmetry for open strings,''
Phys.\ Rev.\ D {\bf 62}, 126005 (2000)
arXiv:hep-th/0006047.
%%CITATION = HEP-TH 0006047;%%
}

%\GovindarajanEF
\lref\GovindarajanEF{
S.~Govindarajan, T.~Jayaraman and T.~Sarkar,
``On D-branes from gauged linear sigma models,''
Nucl.\ Phys.\ B {\bf 593}, 155 (2001)
arXiv:hep-th/0007075.
%%CITATION = HEP-TH 0007075;%%
}

%\HellermanCT
\lref\HellermanCT{
S.~Hellerman and J.~McGreevy,
``Linear sigma model toolshed for D-brane physics,''
JHEP {\bf 0110}, 002 (2001)
arXiv:hep-th/0104100.
%%CITATION = HEP-TH 0104100;%%
}

%\GovindarajanKR
\lref\GovindarajanKR{
S.~Govindarajan and T.~Jayaraman,
``Boundary fermions, coherent sheaves and D-branes on Calabi-Yau
manifolds,''
Nucl.\ Phys.\ B {\bf 618}, 50 (2001)
arXiv:hep-th/0104126.
%%CITATION = HEP-TH 0104126;%%
}

%\DistlerYM
\lref\DistlerYM{
J.~Distler, H.~Jockers and H.~J.~Park,
``D-brane monodromies, derived categories and boundary linear sigma
models,''
arXiv:hep-th/0206242.
%%CITATION = HEP-TH 0206242;%%
}

%\KatzGH
\lref\KatzGH{
S.~Katz and E.~Sharpe,
``D-branes, open string vertex operators, and Ext groups,''
arXiv:hep-th/0208104.
%%CITATION = HEP-TH 0208104;%%
}

%\MorrisonYH
\lref\MorrisonYH{
D.~R.~Morrison and M.~Ronen Plesser,
``Towards mirror symmetry as duality 
for two dimensional abelian gauge  theories,''
Nucl.\ Phys.\ Proc.\ Suppl.\  {\bf 46}, 177 (1996)
arXiv:hep-th/9508107.
%%CITATION = HEP-TH 9508107;%%
}

\lref\kronheimer{PB. Kronheimer and H. Nakajima, 
``Yang-Mills instantons on ALE gravitational 
instantons,'' Math. Ann. {\bf 288}(1990)263}

\lref\mooresegal{G. Moore and G. Segal, unpublished. The 
material is available in lecture notes from the April 2002 Clay 
School on Geometry and Physics, Newton Institute,  and 
http://online.kitp.ucsb.edu/online/mp01/moore1.} 

\lref\iaslectures{E. Witten, in {\it Quantum Fields and Strings: A 
Course for Mathematicians}, vol. 2, P. Deligne et. al. eds. 
Amer. Math. Soc. 1999} 

%%%%%%%%%%%%%%%%%%%%%%%%%%%%%%%%%%%%%%%%%%%%%%%%%%%%%%%%%%%%%%%%%%%%%%%%%%
%%%%%%%%%%%%%%%%%%%%%%%%%%%%%%%%%%%%%%%%%%%%%%%%%%%%%%%%%%%%%%%%%%%%%%%%%%
%	The paper starts here
%%%%%%%%%%%%%%%%%%%%%%%%%%%%%%%%%%%%%%%%%%%%%%%%%%%%%%%%%%%%%%%%%%%%%%%%%%
%%%%%%%%%%%%%%%%%%%%%%%%%%%%%%%%%%%%%%%%%%%%%%%%%%%%%%%%%%%%%%%%%%%%%%%%%%

\rightline{EFI-02-48}
\rightline{RUNHETC-2002-49}
\Title{
\rightline{hep-th/0212059}}
{\vbox{\centerline{On Decay of K-theory}}}
\bigskip
\centerline{Emil J. Martinec$^1$ and Gregory Moore$^2$}
\bigskip
\centerline{$^1$ {\it Enrico Fermi Inst. and Dept. of Physics,
University of Chicago}}
\centerline{\it 5640 S. Ellis Ave., Chicago, IL 60637-1433, USA}
\bigskip
\centerline{$^2$ {\it Department of Physics, Rutgers University}}
\centerline{\it Piscataway, NJ 08855-0849, USA}    

\bigskip
\noindent
Closed string tachyon condensation resolves
the singularities of nonsupersymmetric orbifolds,
however the resolved space typically has fewer
D-brane charges than that of the orbifold.
The description of the tachyon condensation process via a 
gauged linear sigma model enables one to track the topology
as one passes from the sigma model's ``orbifold phase'' to its
resolved, ``geometric phase,'' and thus to follow how the
D-brane charges disappear from the effective spacetime 
dynamics.  As a mathematical consequence, our results 
point the way to a formulation of a ``quantum McKay 
correspondence'' for the resolution of toric orbifold 
singularities. 
 
\vfill
 
\Date{December 4, 2002}

%%%%%%%%%%%%%%%%%%%%%%%%%%%%%%%%%%%%%%%%%%%%%%%%%%%%%%%%%%%%%%%%%%%%%%%%%%
%%%%%%%%%%%%%%%%%%%%%%%%%%%%%%%%%%%%%%%%%%%%%%%%%%%%%%%%%%%%%%%%%%%%%%%%%%

\newsec{Introduction and summary}

The study of tachyon condensation on unstable localized defects
such as D-branes, NS5-branes, and orbifolds
has yielded a number of insights into the structure
of string theory. For example, open string tachyon 
condensation provides one route to  
the topological classification of D-branes via K-theory. 
The decays of  localized defects via closed string tachyon condensation
have exhibited striking analogies to the open string case 
\refs{\AdamsSV,\HarveyWM,\VafaRA,\MartinecTZ}.
There is, however, one notable difference. 
In open string tachyon condensation, the charges of D-branes 
remain invariant.  However, in the typical closed string 
tachyon condensation described in the above references, 
it seems that D-brane charge ``disappears.''
The present paper is an attempt to understand 
in more precise terms how it disappears, 
and where it goes. 

Our central tool for answering the question will be the 
worldsheet renormalization group (RG). 
The worldsheet RG has proven to be 
a reliable tool in analyzing the possible decays and their endpoints
under tachyon condensation for both open and closed strings
(see \MartinecTZ\ for a review).
Nonconformal backgrounds in the worldsheet
field theory provide a way of continuing
off-shell; RG flows interpolate between classical
solutions, and thus provide information about the effective action 
and the topology of the configuration space. 
Tachyon condensation corresponds to adding a relevant
operator to the worldsheet Lagrangian describing the
background in which perturbative strings propagate;
the endpoint of tachyon
condensation in this context is the IR fixed point of the
worldsheet renormalization group flow.

Typically, it is difficult to follow the renormalization
group trajectory of a generic perturbation of the UV
fixed point all the way to its far IR limit;
nonperturbative information is required.  Such information
is provided by the chiral ring (the BPS states) of $\CN=2$
extended worldsheet supersymmetric theories 
\LercheUY; the renormalization of such states
is under good control and enables one to understand
the structure of flows preserving the $\CN=2$ structure.

In fact, in $\CN=2$ conformal field theories, there
are two rings, due to the independent left- and right-moving
supersymmetry algebras: The {\it chiral} ring, consisting of operators
that are left-chiral and right-chiral; and the
{\it twisted chiral} ring, whose operators are left-chiral
and right-anti-chiral.  One can preserve one or the other
but not both along RG flows. 

Orbifolds $\IC^d/\Gamma$, where $\Gamma\subset U(d)$,
provide a large class of examples.  In the present work,
we will mostly consider the abelian orbifolds $\IC^2/\IZ_{n(p)}$
defined by the discrete group action on $\IC^2$
\eqn\dscgpact{
  (X,Y)\longrightarrow (\omega X,\omega^{p}Y)
}
where $\omega=\exp[2\pi i/n]$.  
Note that for $p=n-1$,
the rotation is in $SU(2)$ rather than $U(2)$
so that spacetime supersymmetry is preserved;
these are the well-known $A_{n-1}$ ALE orbifolds.
 
The orbifold twisted chiral ring is built out of the 
$\IZ_n$ twist operators 
$\CT_\kappa$, $\kappa=1,...,n-1$ 
for each separate complex plane:
\eqn\znptwist{
  \CT_\kappa=\CT_{\kappa/n}^{\sst(X)}\CT_{\{\kappa p/n\}}^{\sst(Y)}\ ,
}
where $\{\xi\}$ denotes the fractional part of $\xi$,
$0\le\{\xi\}<1$.
We concentrate here on the GSO projection corresponding
to type 0 strings, for which all of these operators
are present in the string spectrum.%
\foot{The bulk tachyon will be fine-tuned to zero.
We will comment on the type II theory in section 8.}
The operators \znptwist\ carry the $U(1)_\X \times U(1)_\Y $ $R$-charges
\eqn\Rchge{
 % {1\over n} (q_\kappa,p_\kappa)=
	(\kappa/n,\{\kappa p/n\})
}
corresponding to charges under axial
rotations of the $X$ and $Y$ planes, respectively
(for a more detailed discussion, see
\refs{\HarveyWM,\MartinecTZ}).
A plot of these charges for the twisted chiral ring of the $n(p)=10(3)$
orbifold is shown in figure \ring.

\bigskip
{\vbox{{\epsfxsize=2.1in
        \nobreak
    \centerline{\epsfbox{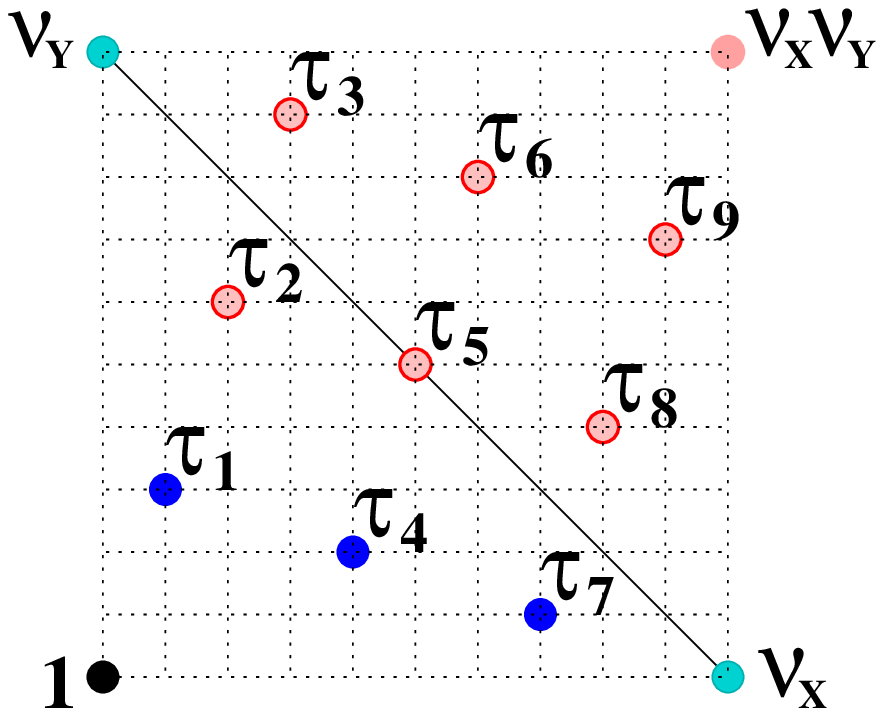}}
        \nobreak\bigskip
    {\raggedright\it \vbox{
{\bf Figure \ring.}
{\it
R-charge vectors $\coeff1n\vv_\kappa =(\kappa/n ,\{\kappa p/n\})$
of twist fields $\CT_\kappa $, 
$\kappa =1,...,9$,
for $n(p)=10(3)$, together with the (untwisted sector) twisted
chiral fields $\CV_{\rm\sst X}$, $\CV_{\rm\sst Y}$
representing the volume forms on the $X$, $Y$ complex planes.
The solid blue dots indicate the generators of the chiral ring.
} }}}}
    \bigskip}

The BPS property determines the total (left plus right)
conformal scaling dimensions
\eqn\scaledim{
  \Delta_\kappa=
%h_\kappa+\bar h_\kappa =
%{(q_\kappa +p_\kappa )\over n} 
	\frac\kappa n+\left\{\frac{\kappa p}{n}\right\}
}
of the twist operators $\CT_\kappa $.  
Thus the chiral operators below the diagonal
line in figure \ring\ are relevant operators corresponding to
closed string tachyons; those on the line generate
marginal deformations, and those above it yield irrelevant perturbations.
Since the twist operators are the only relevant deformations
that we will be considering, 
we will henceforth abuse language and
drop the modifier `twisted'
when referring to the twisted chiral ring, and simply call it 
the chiral ring.  Renormalization group flows
generated by a single relevant scaling operator
were considered in 
\refs{\AdamsSV,\HarveyWM,\VafaRA,\MartinecTZ}.
Below, in section 6, 
we will present a picture of the generic RG flow.

For the supersymmetric $\IC^2/\IZ_{n(n-1)}$ orbifolds,
it is well-known that the generic deformation of the orbifold CFT by 
the $n-1$ marginal twist fields resolves the orbifold singularity,
yielding a nonsingular $A_{n-1}$ ALE manifold.
The algebro-geometric procedure for resolving the singularity
consists of excising the singular point at the origin by
blowing it up into a chain of $n-1$ $\IP^1$'s intersecting
in a pattern specified by the $A_{n-1}$ Dynkin diagram.
For the general non-supersymmetric $\IC^2/\IZ_{n(p)}$ orbifolds,
there is a similar resolution of the singularity known as the 
Hirzebruch-Jung (HJ) or minimal resolution 
\refs{\bpv,\fulton,\riemenschneider}.
This resolution consists of excising the orbifold point
and inserting a chain of $r$ $\IP^1$'s, where $r$ is the number
of terms in the {\it continued fraction expansion} of $n/p$,
\eqn\contfrac{
  \frac np=~a_1-\frac{1}{~a_2-\frac{1}{~a_3-\frac{1}{~\cdots ~1/a_r}}}
	:=[a_1, \dots, a_r] \ ,
} 
for integers $a_\alpha\ge 2$.  This resolution describes
the geometry of the orbifold theory perturbed in a generic
way by the twist operators that generate the chiral ring.
The resolution is called ``minimal'' because there are other 
resolutions of the singularity with more $\IP^1$'s
in the resolution chain.  These are associated to
continued fractions \contfrac\ having some of
the $a_{\alpha}=1$.
More details of these resolutions will be given below in sections 
four and five.

Let us now consider D-branes in these models. 
An orbifold conformal field theory admits a canonical set of
`fractional' D-branes \refs{\DouglasSW,\DiaconescuBR}.
For any representation of $\IZ_n$ there is a corresponding
fractional brane.  
These branes carry charges that couple to 
corresponding RR gauge fields.
The fractional brane charges generate the entire lattice 
of possible D-brane charges.  Mathematically, these
objects generate the equivariant K-theory of the orbifold 
\refs{\WittenCD,\GarciaCompeanRG}.
For the supersymmetric $\IC^2/\IZ_{n(n-1)}$ orbifolds,
this agrees nicely with the compact K-theory 
of the resolved $A_{n-1}$ ALE space: 
There are $(n-1)$ classes corresponding to line 
bundles on the resolving spheres, together with 
the D0 brane.%
\foot{There is a perfect pairing of the $K$-theory 
with the compact $K$-theory. The $K$-theory has a 
natural basis of $n$ distinct canonical line bundles 
(including the trivial bundle)
on the smooth space corresponding to 
space-filling D4-branes with magnetic monopoles threading the various 
$\IP^1$'s of the resolution.} 
This pleasant correspondence, whose 
physical realization is so natural, is part of the story of the 
{\it McKay correspondence} (For a sampling of 
references in the math literature, see  \refs{\reid,\BatyrevJU,
\ItoZX,\itonak,\crawthesis} and in the physics 
literature, see \refs{
\AspinwallXS,\DiaconescuEC,\TakayanagiXT,\MayrAS,\LercheVJ,
\DelaOssaXK,\GovindarajanVI, \TomasielloYM,\HeCR} ).

When we consider the orbifolds $\IC^2/\IZ_{n(p)}$ things are not 
so simple.  
For general $\IZ_{n(p)}$, it will still be the case that
the equivariant K-theory of the orbifold is the representation
ring of $\IZ_n$, so that the lattice of orbifold D-brane
charges is isomorphic to $\IZ^n$. 
That is, there are still $n$ distinct kinds
of compactly supported  D-brane at the orbifold point. 
However, in general the K-theory lattice of D-brane
charges of the smooth Hirzebruch-Jung space
has a rank smaller than $n$.
For $p\ne n-1$ one has $r<n-1$ in \contfrac, 
so that there are $r<n-1$ $\IP^1$'s needed
to smooth the orbifold singularity, and thus, 
taking into account the D0 brane, $r+1<n$ 
generators  
of the compactly supported K-theory lattice of the resolved space.  
Simply put, there are not enough cycles
in the resolved space to wrap D-branes on 
to account for all the independent D-brane charges
of the orbifold one started with.
{\it Where did the extra D-brane charges go?}
One of the main goals of this paper is to provide an answer to
this question.

In order to answer this question it is very 
convenient to introduce a gauged linear sigma model (GLSM)
\WittenYC,
for which the UV fixed point is the orbifold 
conformal field theory. 
The GLSM construction also contains couplings
to twisted sector tachyons which resolve the orbifold singularity,
realized as Fayet-Iliopoulos 
parameters of its abelian gauge dynamics. 
Some aspects of this type of GLSM were studied in \VafaRA.

The GLSM consists of a $U(1)^r$ gauge group 
coupled to charged chiral matter fields.
It has both Coulomb and Higgs branches
of its configuration space. 
In the IR, the Higgs branch can be 
interpreted as a nonlinear sigma model whose target space 
is a resolution of $\IC^2/\IZ_{n(p)}$,%
\foot{Naturally associated to the description of
the singularity and its resolution in toric geometry, \cf\ \fulton.}
such as the Hirzebruch-Jung resolution.  
Of course, along the RG flow
one has a massive 2D quantum field theory, 
and the data of the closed string geometry
are undergoing RG flow. 
However, at a fixed RG scale one can speak of the D-branes 
in the massive Higgs branch theory. 
In the IR where the target space of the Higgs branch is a smooth manifold,
these D-branes have a geometrical interpretion 
as branes wrapping nontrivial cycles of the smooth Hirzebruch-Jung manifold; 
they are therefore interpreted as K-theory classes 
of the resolved space. 
These D-brane charges are the $r+1$ ``obvious'' charges.
%(plus one more trivial charge corresponding to pointlike
%\eg\ D0-branes on the Hirzebruch-Jung space). 

We will also find $n-r-1$ independent additional D-branes living on
the Coulomb branch of the configuration space,
one for each of the distinct massive vacua on this branch. 
In section 7.1 below we will show that these objects
supply the extra charges needed to account fully 
for all the D-brane charge present in the orbifold.  
This is the resolution of our puzzle. 

There are three interesting byproducts of this result.

The first byproduct is an improved understanding of the 
endpoint of the generic RG flow associated to closed 
string tachyon condensation of the $\IC^2/\IZ_{n(p)}$ 
orbifold, extending the results of 
\refs{\AdamsSV,\HarveyWM,\VafaRA,\MartinecTZ}. 
The general picture is one of several 
ALE spaces separating from one another along the flow. 
The precise pattern of ALE spaces 
is encoded in the minimal continued 
fraction expansion \contfrac. 
To each consecutive subsequence 
$[a_{\alpha},...,a_{\alpha+\ell}]$ of \contfrac\
having all $\ell$ of the $a$'s equal to two,
there is an $A_{\ell}$ ALE space 
comprising one component of the IR limit of the geometry.
The various ALE components are separated by an infinite
distance as one flows to the IR.
This picture is explained in section 6 below. 

The second byproduct is a suggestion for 
the form of the effective action of the
RR gauge fields that couple to the corresponding
decoupling D-branes; this effective action appears to
take the same form as one finds 
in the analogous open string examples:
\eqn\RReffact{
  \CS^{\sst RR}_\eff=\int d^6 \! x f(T)[F_{RR}^2+\ldots]\ ,
}
where   $f(T)\to 0$ as the tachyon condenses.
The RG approach thus appears to support the idea
that the decoupling of effective fields and charges
under tachyon condensation takes a universal form
for both open and closed string degrees of freedom.%
%\foot{It would be interesting if closed string field theory
%could supply an independent check of this conjecture.}
%

The third byproduct is an application to mathematics, 
where our construction suggests 
a generalization of the McKay correspondence 
mentioned above.
A generalization of the McKay correspondence to 
orbifolds of the type $\IC^2/\IZ_{n(p)}$ has been discussed in 
the mathematical literature
\refs{\riemenschneider,\wunram,\ishii,\ito}.
These authors compared the equivariant K-theory of the 
orbifold with the compactly supported K-theory of the 
minimal resolution.  
In physical terms, they were able to characterize 
``special representations'' of the quantum $\IZ_n$ 
symmetry of the orbifold corresponding to 
D-branes on the Hirzebruch-Jung resolution in such a way as to 
map the K-theory of the former onto the latter 
(the isomorphism even extends to the derived categories). 
In section 5 we show that the 
GLSM point of view makes the construction of these 
``special representations'' very natural. The 
K-theory of the minimal resolution is generated by 
tautological sheaves -- line bundles associated 
to a principal $U(1)^r$ bundle over the resolution 
via the unitary irreps $\rho_i$ of $U(1)^r$. 
We show that at the orbifold point the gauge group $U(1)^r$ is 
spontaneously broken to a discrete $\IZ_n$ subgroup, 
and that when restricted to $\IZ_n$, 
the representations $\rho_i$ are precisely 
the ``special representations'' of 
\refs{\riemenschneider,\wunram,\ishii,\ito}.
We expect that our  construction will apply to  
all orbifolds that admit a toric resolution, thus allowing 
a construction of such ``special representations'' in 
a large class of examples. 
More interestingly, the resolution of our paradox suggests 
that, by adjoining the  charges of the 
D-branes of the  massive vacua on the 
Coulomb branch of the GLSM to the K-theory of the resolved space, there 
should be a ``quantum McKay correspondence'' that holds 
for a large class of orbifold singularities that admit 
toric resolutions.

%%%%%%%%%%%%%%%%%%%%%%%%%%%%%%%%%%%%%%%%%%%%%%%%%%%%%%%%%%%%%%%%%%%%%%%%%%

\newsec{The Gauged Linear Sigma Model}

The orbifold spacetimes we consider are toric varieties, 
that is, they can be described as quotients by a $U(1)^r$ action.  
A simple way of generating such a quotient is to employ
a gauged linear sigma model (GLSM). In this section we will 
recall some of the standard facts which will be important in 
what follows. All of the results can be found in 
\refs{\WittenYC,\MorrisonFR,\iaslectures,\HoriKT.}. 
For brief summaries, see \refs{\MorrisonYH,\HoriFJ}.

Complex target space geometry and worldsheet supersymmetry 
imply $\CN=2$ worldsheet supersymmetry.
Therefore, consider $r$ abelian $\CN=2$ gauge fields
$V_\alpha$, $\alpha=1,...,r$ coupled to $r+d$ 
$\CN=2$ chiral matter fields
$X_i$ with charges $Q_{\alpha i}$.  
The field strengths of the gauge fields 
are contained in twisted chiral superfields%
\foot{Here the bar denotes worldsheet complex conjugation, 
and the star denotes complex conjugation in field space. 
We will usually follow the conventions of Hori and Vafa \HoriKT.}  
$\Sigma={1\over 2} \{\overline\CD,\CD^*\}$.
The classical Lagrangian is
\eqn\glsmact{
  \CL = \int d^4\theta \;\left(\Xbar_i e^{2Q_{\alpha i} V_\alpha} X_i
	-\frac{1}{2e_{\alpha}^2}\bar\Sigma_\alpha \Sigma_\alpha\right)
	-\half\left(\int d^2\tilde\theta 
		\;t_{\alpha}\Sigma_\alpha+{\rm c.c.}\right)\ ,
}
%%%%
%%%%
where repeated indices are summed and 
\eqn\teebare{
t_{\alpha } =\zeta_\alpha-i\theta_\alpha
}
 combines
the Fayet-Iliopoulos (FI) parameter $\zeta$ and theta angle $\theta$
for the $\alpha^{\rm th}$ gauge field; $d^2\tilde\theta$ is
the twisted chiral superspace measure.  

In order to define the quantum theory we must
renormalize the theory. Accordingly we 
introduce a momentum cutoff $\Lambda$ and  fix a 
renormalization scale $\mu$. 
The   1-loop renormalization of 
the FI parameters is 
\eqn\FIren{
  t_{\alpha,{\rm eff}}(\mu) = t_{\alpha,{\rm bare}} 
	+ \sum_{i=1}^{r+d} Q_{\alpha i} \log{\mu\over \Lambda} 
}
where  $t_{\alpha,{\rm bare}} $ are bare parameters 
defined at the momentum cutoff scale $\Lambda$. 
Note that the theory also has dimension one couplings 
$e_{\alpha}$.   The renormalized theory is 
defined by taking $\Lambda \to +\infty$ holding 
$t_{\alpha,{\rm eff}}(\mu),\mu$ and $e_{\alpha}$ 
fixed. The scale dependence of couplings depends 
crucially on the sign of the beta function, which 
is governed by: 
\eqn\betafun{
b_{\alpha} := \sum_i Q_{\alpha i}
}
Note that this requires that we take 
$t_{\alpha,{\rm bare}} \to -\infty$ if $b_{\alpha} <0$
and 
$t_{\alpha,{\rm bare}} \to +\infty$ if $b_{\alpha} >0$.

Our general strategy will be to use the GLSM to define a model which, 
at a high energy scale, say $\mu \sim \Lambda$,  
is close to the $\IC^2/\IZ_{n(p)}$ orbifold CFT fixed point, 
and whose RG flow is ``close'' to that of the orbifold CFT perturbed by 
relevant operators from the chiral ring, see figure \glsm.  
We are then interested in 
the low energy behavior of the theory, that is, in the IR limit of the 
RG flow of such theories.  This should be a good approximation to the 
RG flow of the perturbed orbifold theory.  Therefore, we now turn to 
a discussion of the low energy physics of the GLSM.

 \bigskip
{\vbox{{\epsfxsize=2.2in
        \nobreak
    \centerline{\epsfbox{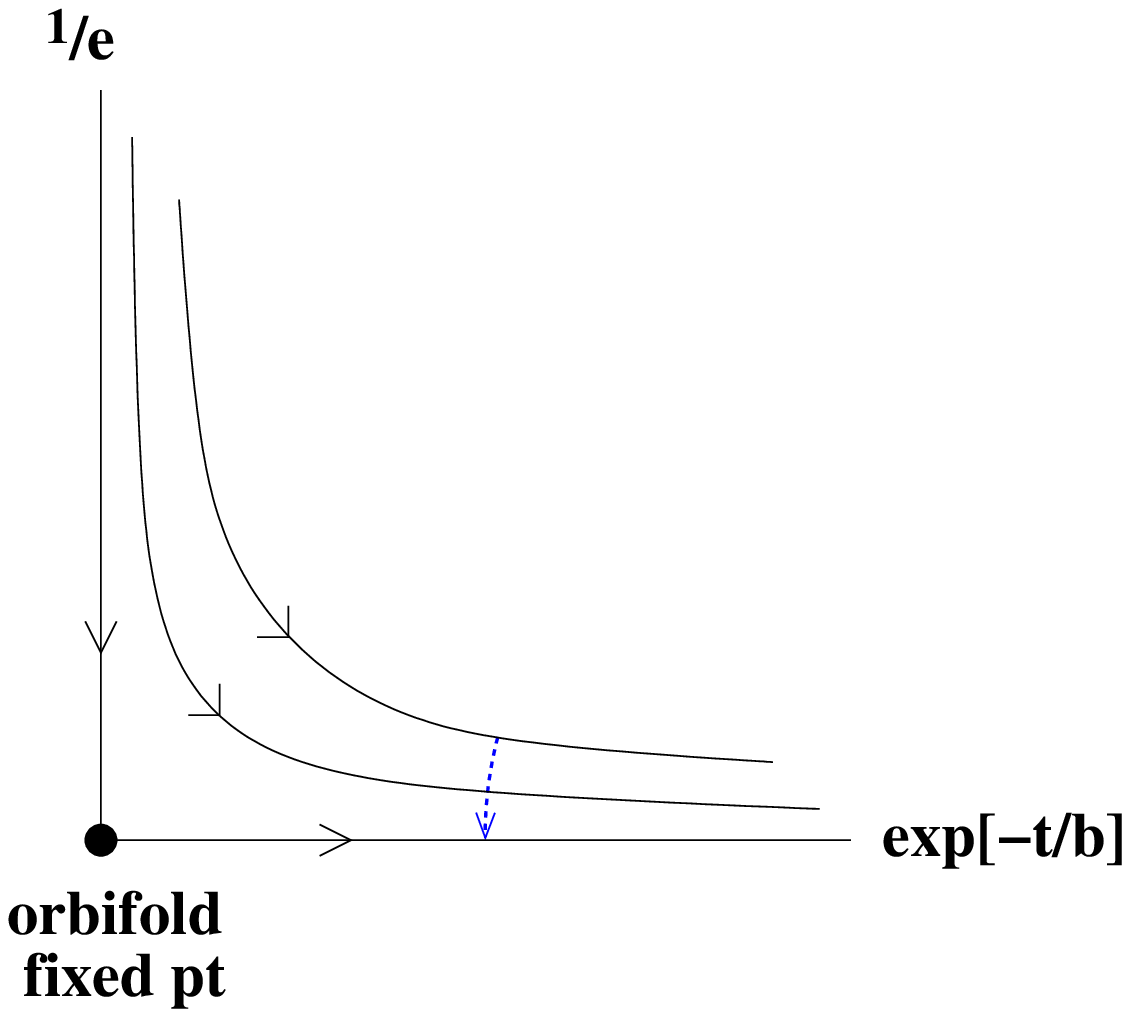}}
        \nobreak\bigskip
    {\raggedright\it \vbox{
{\bf Figure \glsm.}
{\it
Schematic description of the renormalization
group trajectories for the couplings 
$(\exp[-t_\alpha(\mu)/b_\alpha ], 1/e_{\alpha})$.
The RG fixed point at the origin is the orbifold CFT. 
The flow out of the fixed point along the horizontal axis 
is the RG flow of the orbifold perturbed by relevant chiral operators. 
This flow is the limit (indicated by the dashed blue line)
of flows defined by choosing a fiducial scale
$\mu_*$ and sending $e$ at that scale to $\infty$. } }}}}
    \bigskip}

To determine the low energy behavior we must examine the potential 
energy for the fields. In the classical theory the potential 
energy takes the form: 
\eqn\Xpotl{
  U_{\rm classical}=\sum_{\alpha=1}^r
	\frac{e_\alpha^2}{2}\bigl(M_\alpha(X)-\zeta_\alpha\bigr)^2
	+\sum_{\alpha,\beta=1}^r\bar\sigma_\alpha\sigma_\beta
		\sum_{i=1}^{r+d}Q_{\alpha i}Q_{\beta i}|X_i|^2 
}
In the second term $\sigma_{\alpha}=\Sigma_{\alpha}
\vert_{\tilde \theta=0} $. 
The first term comes from solving the equation of motion for the auxiliary
field $D_\alpha$ in the vector multiplet, where we have defined
\eqn\Dtermsmm{
  M_\alpha(X) := \sum_i Q_{\alpha i} |X_i|^2\ .
}
 
The classical ground states are easily determined:  
Both terms in \Xpotl\ are positive semidefinite and there are in general
two branches of solutions.  The first, the 
{\it Higgs branch}, has $\sigma_\alpha=0$ and 
nonvanishing values of $X_i$. In general, nonvanishing 
values of $X_i$ transform nontrivially under $U(1)^r$, that is, 
these classical VEV's 
break the $U(1)^r$ gauge symmetry, 
hence the name ``Higgs branch.''  The Higgs branch 
is described by solving:  
\eqn\Dterms{
   \sum_i Q_{\alpha i} |X_i|^2
= \zeta_\alpha .
}
Let us denote the solution set to  \Dterms\  by
$\CS_{\vec\zeta}\subset \IC^{r+d}$. Taking into account 
gauge invariance, we see that the classical Higgs branch of vacua  
is the set $\CS_{\vec\zeta}/U(1)^r$.

Mathematically, one can view the functions $M_\alpha(X)$
as Hamiltonian functions on a phase space 
whose symplectic form is the \kahler\ form 
$\Omega={i\over 2}  dX_i \wedge  d\Xbar_i$.  Then the conditions \Dterms\
(called the {\it moment map equations})
fix a level set of $M_\alpha$, and the quotient by the
$U(1)^r$ torus action generated by the $M_\alpha$
results in a Hamiltonian reduction
of the phase space to $\CS_{\vec\zeta}/U(1)^r$.  
In this setting, the reduction is known as a {\it \kahler\ quotient}.

If the solution space $\CS_{\zeta}$ admits $X_i=0$ for some $i$ 
then there can be another 
branch of classical vacua, the {\it Coulomb branch}, 
where some subgroup of $U(1)^r$ 
is unbroken, and some $\sigma_{\alpha}$ can take (continuous) 
nonzero expectation 
values.  However, for generic values of $\zeta_{\alpha}$ such branches 
are absent. 

Now let us turn to the quantum mechanical theory renormalized as in 
\FIren. There are still 
Higgs and Coulomb branches, but there are also  
several important modifications of the above description of the 
space of vacua.  

The Higgs branch equations \Dterms\ are 
modified by setting $M_{\alpha}(X) = \zeta_{\alpha,{\rm eff}}(\mu)$. 
The IR physics is determined by the behavior of $\CS_{\zeta(\mu)}/U(1)^r$
for   $\mu \to 0$. As we will see in sections 3,4,5 
the nature of the model depends very strongly on the sign 
of $b_{\alpha}$.  In the case where the space 
$\CS_{\vec \zeta(\mu)}/U(1)^r$ is a smooth manifold with 
$c_1(TX)<0$ the low energy dynamics is that of a nonlinear 
sigma model with target space  
$\CS_{\vec \zeta(\mu)}/U(1)^r$.  
\foot{See a related discussion in 
\iaslectures\ where the RG 
flow on the Higgs branch is described as a flow from 
a LG model to a nonlinear model with $c_1(TX)<0$ or a 
flow from a nonlinear model with $c_1(TX)>0$ to a LG model, 
depending on the sign of $b_{\alpha}$. }

Since the sigma model metric is expected to be renormalized,
the K\"ahler quotient metric will not be precisely that of
the renormalized theory at a finite scale $\mu$. 
Nevertheless, we will use the K\"ahler quotient metric
as a {\it qualitative} guide to the geometry.
 
The most dramatic modification of the configuration space
of the quantum theory takes place
on the Coulomb branch. 
We will assume that $b_{\alpha}\not=0$ in what follows. 
% 
%
%In addition, so long as the beta function(s) are nonvanishing, there 
%are also more subtle {\it Coulomb branch} vacua, whose existence 
%is purely quantum mechanical. 
%
In the classical theory the Coulomb branch is either absent, or is
a continuous manifold.  In the quantum theory, the Coulomb branch
is a discrete set of vacua, which are massive 
if the kinetic term for $\sigma_{\alpha}$ is nonsingular.
These vacua are essential in the 
quantum McKay correspondence, so let us recall how they arise. 

The Coulomb branch vacua arise (for $\sum Q_{\alpha i} \not=0$) 
when the $\sigma_\alpha$ gain vacuum expectation values,
giving mass to all the $X_i$.
The $X_i$ fields can then be integrated out; 
doing so results in 
an effective twisted superpotential: 
\eqn\renspotl{
  \widetilde W_\eff = 
	- \sum_{\alpha=1}^r \Sigma_\alpha \left(t_{\alpha,{\rm eff}}(\mu)
	+ \sum_{i=1}^{r+d} Q_{\alpha i}\log\Bigl({1\over e\mu}
		\sum_{\beta=1}^r Q_{\beta i}\Sigma_\beta \Bigr)\right)
}
($e$ here -- and here only -- is the transcendental number $2.71...$, 
not the gauge coupling)
which neatly summarizes the running of the couplings. 
By standard holomorphy arguments, 
this formula is 
exact \refs{\iaslectures,\HoriKT}. 
The bosonic effective
potential of the twisted scalars $\sigma_\alpha$ 
deduced from $\widetilde W_\eff$ is: 
\eqn\sigpotl{
  \tilde U(\sigma) =  \sum_{\alpha=1}^r \frac{e_{\alpha}^2}2
	\left\vert \;   t_{\alpha,{\rm eff}}(\mu)
        + \sum_{i=1}^{r+d} Q_{\alpha i}\log\Bigl({1\over \mu}
                \sum_{\beta=1}^r Q_{\beta i}\sigma_\beta \Bigr)\right\vert^2
}
Integrating out the $X_i$ fields is justified at momentum scales
below the scale set by their masses; 
this is determined by the VEV's of the $\sigma_\alpha$, namely
\eqn\sigvev{
  \vev{\sigma_\alpha}\sim 
	\mu\,\exp[-t_{\alpha,{\rm eff}}(\mu)/b_\alpha ]\ .
}
Because of the effects of pair creation on the vacuum,
we must minimize over the 
branches of $\theta_{\alpha} \to \theta_{\alpha} + 2 \pi n_{\alpha}$. 
This renders the potential \sigpotl\ single-valued  in $\sigma_{\alpha}$.
Generically, this potential has several local minima,
which are the vacua of the Coulomb branch.
The kinetic terms of the 
$\sigma$ fields likewise are expected to receive renormalization. 
These effects can be very important, for example, in 5-brane 
physics, but we believe that they are not important in the 
examples studied in this paper. It would be good to clarify this 
point. 

Finally, we note that in order to 
reproduce the physics of the orbifold CFT in the UV, one must 
consider the limit $e_{\alpha}^2 \to \infty$ 
of the renormalized theory.
The renormalization group trajectories 
take the schematic form of figure \glsm\ above, and only approach 
the  RG trajectory of the perturbed orbifold in this limit. 
We should make sure that $e_{\alpha} \to \infty$ does not violate any 
of the key assumptions made above;
in particular, one should check the self-consistency of
the effective action \sigpotl\ that leads to the Coulomb
branch vacua.  The limit $e_{\alpha}^2 \to \infty$ 
leaves fixed the scales \sigvev\ set by the VEV's of the $\sigma_\alpha$.
The dimensionless scalars
$\hat \sigma_{\alpha}= \sigma_{\alpha}/e_{\alpha}$
with canonical kinetic terms 
have masses $\tilde U'' \sim e_{\alpha}^4/\vev{\sigma_\alpha}^2$.
Thus the $\sigma_\alpha$ fluctuations about the
Coulomb branch minima \sigvev\ become very heavy in this limit
and decouple.
The end result is that the  
Coulomb branch vacua are well-separated from the Higgs branch 
vacuum by a large potential barrier. 
This is the essential fact we will need.

%%%%%%%%%%%%%%%%%%%%%%%%%%%%%%%%%%%%%%%%%%%%%%%%%%%%%%%%%%%%%%%%%%%%%%%%%%

\newsec{Warmup: $\IC^2/\IZ_{a(1)}$} 

Perhaps the simplest class of examples of the phenomenon
of ``disappearing'' topology are the orbifolds $\IC^2/\IZ_{a(1)}$.
In this section, we will review how this space and
its resolution are described as different ``phases''
of the GLSM, and give the prototype of the
resolution of the puzzle stated in the introduction.

Consider the $U(1)$ \kahler\ quotient on  
$(X_0, X_1, X_2)$ with charges $(1,-a,1)$ 
where $a$ is a positive integer.%
\foot{In complex dimension $d=2$, it will prove convenient
to shift the range of the index on the chiral field by one:
$i=0,...,r+1$ rather than $1,...,r+2$.  We will do so in
the remainder of the discussion.} 
There is a single D-term equation \Dterms
\eqn\mm{
\mu:= \vert X_0 \vert^2 - a \vert X_1 \vert^2 + \vert X_2 \vert^2 = \zeta 
}
which has a solution set $\CS_{\zeta} \subset \IC^3$. 
We want to describe the geometry of the quotient $\CS_\zeta/U(1)$
which arises when we fix the action of the gauge symmetry on $\CS_{\zeta}$. 
This depends strongly on the sign of $\zeta$;
for $\zeta<0$ one finds the $\IZ_{a(1)}$ orbifold
singularity at the origin, and for $\zeta>0$
one finds a smooth resolution of the space. 

%%%%%%%%%%%%%%%%%%%%%%%%%%%%%%%%%%%%%%%%%%%%%%%%%%%%%%%%%%%%%%%

\subsec{Description of the quotient space: $\zeta< 0 $} 

For $\zeta < 0$ the variable $X_1$ is necessarily 
nonzero and can be used to fix a gauge (up to a discrete quotient): 
\eqn\neighb{
(X_0, X_1, X_2) = (e^{i \theta} \xi_1, 
	e^{-i a \theta} \vert X_1 \vert , e^{i\theta} \xi_2)
}
where $\xi_1, \xi_2\in \IC^2$.
Choosing this positive root means that 
$(\xi_1, \xi_2)$ are gauge invariant up to multiplication 
by $\omega\in \IZ_a$, 
\eqn\orbact{
  (\xi_1, \xi_2) \sim (\omega \xi_1, \omega \xi_2) 
	\quad,\qquad  \omega\in \IZ_a\ .
}
Plugging \neighb\ into the GLSM action \glsmact,
as $\zeta\to-\infty$ the D-term potential freezes the value
of $|X_1|\to\infty$, and $\xi_{1,2}$ become free fields.
The residual discrete $\IZ_a$ gauge action results
in the orbifold space $\IC^2/\IZ_{a(1)}$.

For finite $\zeta<0$,
the level set $\mu=\zeta$ is simply the subspace
\eqn\mag{
\vert X_1 \vert  = + {1\over \sqrt{a}} \sqrt{\rho_1^2 + \rho_2^2 -\zeta } 
}
where we introduce the magnitude and phase
\eqn\magphse{
\xi_i := \rho_i e^{i \gamma_i} \qquad i =1,2\ .
}
Note that since $\zeta<0$ the argument of the square root in 
\mag\ is always positive.
Thus, restricting to fixed value of $\vert X_1\vert $, the 
quotient space is a Lens space $L(a,1)=S^3/\IZ_{a(1)}$.
Roughly speaking, as a space, $\CS$ is a $U(1)$ bundle over the orbifold 
$\IC^2/\IZ_{a(1)}$. This is not completely accurate 
because at $\xi_1=\xi_2=0$ 
the orbifold group acts nontrivially on the fiber. 

%%%%%%%%%%%%%%%%%%%%%%%%%%%%%%%%%%%%%%%%%%%%%%%%%%%%%%%%%%%%%%%

\subsec{Description of the quotient for $\zeta >0$}

For $\zeta > 0$, the choice \mag\ still solves the 
D-term equations \mm, provided we satisfy the
inequality $\rho_1^2 + \rho_2^2 \geq  \zeta$;
however, the locus $\xi_1=\xi_2=0$ -- which resulted in a singular
$U(1)$ quotient for $\zeta<0$ -- is lifted 
from the effective field space for $\zeta>0$.  
If we consider the minimal 3-sphere 
\eqn\threesphere{
\vert \xi_1 \vert^2 + \vert \xi_2\vert^2  = \zeta 
}
in $\CS_{\zeta}$ then $X_1=0$ and the $U(1)$ group is 
still completely unbroken by $X_1$. Therefore in the 
quotient $\CS_{\zeta}/U(1)$ this subset projects to 
$\IP^1$.  In other words,
\eqn\minsthree{
\{ X_1=0 \} \cap \CS_{\zeta} \to \IP^1\ ,
}
which is to be contrasted with the $U(1)$ quotient of the 
subset 
\eqn\xeps{
  \{ X_1 = \epsilon \} \cap \CS_{\zeta} \ ;
}
the fiber of this latter space when projected to the quotient 
is $\IZ_a$, so these subspaces project to Lens spaces $L(a,1)$ 
inside the quotient. 
Thus, the singular set of the $U(1)$ quotient for $\zeta<0$
has been replaced by a nonsingular $\IP^1$ for $\zeta>0$.
That is, the singularity has been resolved.

A special case is $a=1$, where the `orbifold' phase $\zeta<0$
is nonsingular, and the $\zeta\to-\infty$ space is $\IC^2/\IZ_1\equiv\IC^2$.
The regime $\zeta>0$ is however still nontrivial, and describes the 
blowup of $\IC^2$ at a point.  Note in particular that the 
topology of the blown up space is different.

An alternative way to see the geometry for $\zeta>0$
uses the complex geometry of the quotient space.
For $\zeta>0$, either $X_0\not=0$ or $X_2 \not=0$ 
for every point on the solution set $\CS$. Therefore, 
we divide the solution set into two patches, $X_2 \not=0$ and 
$X_0\not=0$, and introduce gauge invariant {\it holomorphic} coordinates 
\eqn\holocoordplus{
\eqalign{
z_+ & = X_0/X_2   \cr
p_+ & = X_1 X_2^a \cr}
}
on the patch $X_2\not=0$, and 
\eqn\holocoordminus{
\eqalign{
z_- & = X_2/X_0   \cr
p_- & = X_1 X_0^a \cr}
}
on the patch $X_0\not=0$. 
We now recognize that $z_\pm$ are coordinates on $\IP^1$, and 
$p_- = p_+ z_+^a$, so that $p_\pm$ are fiber coordinates on the 
complex line bundle $\pi: \CO(-a) \to \IP^1$. 

Incidentally, the relationship between 
the holomorphic coordinates $(z,p)$ and  the 
coordinates $\xi_i$ defined above is the following. 
On the patch $X_2 \not=0$ the coordinates $(z_+,p_+)$ relate
to the coordinates $(\xi_1, \xi_2)$ as follows:
\eqn\ptoxi{\eqalign{
  z_+ &= \xi_1/\xi_2 \cr
  p_+ &= \xi_2^a \sqrt{(\vert \xi_1\vert^2 + \vert \xi_2\vert^2 - \zeta)/a} 
}}

%{\bf Work out and include something on charges $(-p,n,-q)$. } 

%%%%%%%%%%%%%%%%%%%%%%%%%%%%%%%%%%%%%%%%%%%%%%%%%%%%%%%%%%%%%%%

\subsec{Homology and $K$-theory} 

The homology and cohomology groups are
\eqn\znonehom{\eqalign{
H_j (\CO(-a) )  & =  \cases{\IZ & $j=0$ \cr 
			\IZ & $j=2$ \cr
			0  & $j=4$ } \cr
 & \cr
H^j (\CO(-a) )  & =  \cases{\IZ & $j=0$ \cr 
			\IZ & $j=2$ \cr
			0  &  $j=4$ }\quad .
}}
Poincare duality says that 
$H^k(X) \cong H_{n-k}(X, \p X) \cong H_{n-k, cpt}(X) $ 
and $H^k_{cpt}(X) = H^k(X,\p X) \cong H_{n-k}(X)$. 
The intersection form on $H_2$ is simply $-a$. 

The K-theory is isomorphic to the cohomology 
(as $\IZ$-modules, not as rings), so 
\eqn\OK{\eqalign{
K^0(X) &= \IZ \oplus \IZ \cr
K^0_{cpt}(X) &= \IZ \oplus \IZ \ .
}}
Now let us compare this with 
the equivariant $K$-theory of K\"ahler quotient in 
the $\zeta<0$ phase: 
\eqn\genequK{
  K_G(\IC^2) \cong R(G)
}
since $\IC^2$ is equivariantly contractible
($R(G)$ is the representation ring of $G$). 
For us,
\eqn\znoneK{
  R(\IZ_a) = \IZ[x]/(x^a=1)
}
as a $\IZ$-module this is $\IZ^a$, and 
this seems to have little to do with the $K$-theory of the 
Hirzebruch-Jung resolution space $\CO(-a)$. 
This is an example of our basic paradox. 

%%%%%%%%%%%%%%%%%%%%%%%%%%%%%%%%%%%%%%%%%%%%%%%%%%%%%%%%%%%%%%%

\subsec{ The canonical line bundle and connection} 

In accounting for D-brane charges it is useful to have a clear idea of 
a basis for the K-theory. 
There is a canonical line bundle $\CR \to \CO(-a)$ which is the 
complex line bundle associated to 
the principal $U(1)$ bundle $\CS_{\zeta} \to \CO(-a)$
by the fundamental representation. 
This bundle carries a canonical connection defined by the 1-form on $\IC^3$: 
\eqn\canconn{
\Theta   = {i\over 2\CN  }  \biggl[ \sum_i  Q_i
\bigl( \bar X_i d X_i - X_i d \bar X_i \bigr)  \biggr]\ . 
}
where 
\eqn\cncnnii{
\CN  = \sum_i Q_i^2 \vert X_i\vert^2
}
Restricted to a gauge orbit \neighb\ this connection gives
$ \Theta  = d \theta $;
moreover, the Lie derivative along $d/d\theta$ is zero, 
and so $\Theta$ is indeed a connection. 
As shown in \refs{\HoriCK,\HoriIC}, this gauge field
can be incorporated into an $\CN=2$ supersymmetric
boundary interaction in the GLSM.
The 2-form 
$ d \Theta  \vert_{\CS_{\zeta}}  $
is a basic form for $\zeta>0$, \ie, we can write it as $\pi^*(F )$. 
Restricting $F$ to $p=0$, 
we find $F$ integrates to $2\pi$ on this sphere, 
so that the Chern-class of the line bundle is $+1$. 
The bundle $\CR$ together with the trivial bundle
generate the first line of \OK.

%%%%%%%%%%%%%%%%%%%%%%%%%%%%%%%%%%%%%%%%%%%%%%%%%%%%%%%%%%%%%%%%%%%%%%%%%%

\subsec{Resolution of the puzzle for $\IC^2/\IZ_{a(1)}$}

In this simple case of a single $U(1)$ gauge field,
the minima of the effective potential \sigpotl\ lie at
\eqn\znonemin{\eqalign{
  \sigma_1^{(\ell)}
	&=\Lambda \,c \, 
		\exp\left[\frac{t_{1,\rm bare}+2\pi i\ell}{a-2}\right]
	=\mu\, c \, 
		\exp\left[\frac{t_{1,\eff}(\mu)+2\pi i\ell}{a-2}\right]
	\quad,\qquad \ell=1,...,a-2\ . 
}}
where $c  = \exp[-a\log(-a)/(a-2)]$, and we are assuming $a\not=2$. 
For $a>2$, we see that there are $a-2$ supersymmetric vacua at
large $\sigma$ for large positive $t_{1,\eff}(\mu)$.
%
%Since $Re\left(t_{1,\eff}(\mu)\right)$ 
%indeed grows toward large positive
%values along the RG flow to the IR ($\mu\to 0$) for $a>2$,
%we see that as we pass into the IR, this picture of Coulomb
%branch vacua will be increasingly accurate.
%
For such values of $t_{1,\eff}(\mu)$ the picture of 
Coulomb branch vacua will be accurate. 
D-branes in such massive vacua are localized at
the extrema \znonemin\ of the effective superpotential,%
\foot{This is true in the topologically twisted theory 
simply because in the topological theory every primitive 
idempotent of the Frobenius algebra leads to a single D-brane 
\mooresegal.  An explicit construction of the 
$\CN=2$ Landau-Ginzburg boundary states is 
described in \HoriCK\HoriIC\GovindarajanEF. } 
thus there are $a-2$ independent D-branes
associated to the Coulomb branch vacua;
these, together with the two localized brane charges
of the Higgs branch $\CO(-a)$ geometry, account
for the full rank $a$ of the equivariant K-theory
lattice of D-brane charges of the orbifold.

Of course, for $a=2$ one has the spacetime supersymmetric
$A_1$ ALE space as the resolution of the orbifold.
The coupling $t_1$ is marginal, and there is no Coulomb branch
of the GLSM configuration space.  Of course, we weren't
looking for one, since the two localized brane charges 
of the geometrical resolution of the orbifold account for 
all of the D-brane charges found in the orbifold.

Finally, for $a=1$ one has the blowup of $\IC^2$ discussed 
in subsection 2 above.  In this case, the RG flow runs in
the opposite direction: In the flow to the IR, i.e. 
as  $\mu\to 0$,  the $\IP^1$ of the resolved space
blows down to a point   and disappears, 
so that $\IC^2/\IZ_1=\IC^2$ is the IR limit of the flow;
the UV fixed point is then the sigma model with the $\IP^1$
infinitely blown up.  Thus again the UV fixed point theory has
one more D-brane charge than the IR geometry 
(associated to the $\IP^1$).  This ``missing'' charge is again found on 
the Coulomb branch in the IR theory; it is just that
according to \FIren,
$Re(t_{1,\eff}(\mu)) \to-\infty$ rather than $+\infty$ along the flow,
compatible with the VEV for $\sigma_1$ given by \znonemin, 
increasing as $\mu \to 0$.%
\foot{This example supports the idea of the Coulomb branch
vacua decoupling from the Higgs branch, rather than forming
a throat as in fivebrane physics.  The formation of such a throat
here would imply the existence of a stable object in 
string theory on flat space carrying no conserved charges.} 

In the next few sections, we will generalize this
result to arbitrary nonsupersymmetric $\IC^2/\IZ_{n(p)}$
orbifolds.  We will see that, apart from the additional
complication of multiple $U(1)$ gauge fields, corresponding
to multiple curves in the resolution of the orbifold
singularity, the basic structure is much the same as we 
have just encountered in the rank one case.
In particular, a careful analysis of the Coulomb 
and Higgs branches of the GLSM will account for all of the
D-brane charges of the orbifold CFT.

%%%%%%%%%%%%%%%%%%%%%%%%%%%%%%%%%%%%%%%%%%%%%%%%%%%%%%%%%%%%%%%%%%%%%%%%%%

\newsec{Generalized Cartan matrices and continued fractions} 

The general orbifold $\IC^2/\IZ_{n(p)}$ can be realized
as a quotient of an $(r+2)$-dimensional space 
by a $U(1)^r$ action generalizing the construction of the 
previous section. This allows one to resolve the singularity
in a similar fashion.   
The resolution  
involves a sequence of $r$ blowups, and should therefore 
be realized as a phase in a $U(1)^r$ GLSM.  The algebraic
geometry of the resolution is for instance explained in
\refs{\fulton,\MartinecTZ}.  The sequence of blowups
produces a chain of $\IP^1$'s; the north pole of the $\alpha^{\rm th}$
$\IP^1$ intersects the south pole of the $(\alpha+1)^{\rm st}$ $\IP^1$,
and the self-intersection numbers are $-a_\alpha$, $\alpha=1,...,r$.
Among the toric resolutions of the orbifold there is 
a ``minimal resolution'' of smallest $r$. 
This minimal resolution has all $a_\alpha\ge 2$.

The blown up space for any such resolution
(not just the minimal one)
is covered by $r+1$ coordinate patches $\IC^2$.
On overlaps, the coordinates $u$, $v$ of successive
patches are related via
\eqn\overlaps{
\eqalign{
  v_{\alpha+1} &= u_\alpha^{-1} \cr %\quad,\qquad
	u_{\alpha+1} &=v_\alpha u_\alpha^{a_\alpha}\quad . }
}
Note the appearance of the transition functions
for the patches \holocoordplus, \holocoordminus\ covering
$\CO(-a_\alpha)$, so that indeed the curves have
the advertised self-intersection numbers.
Note also that the coordinate on the normal bundle of
the $\alpha^{\rm th}$ $\IP^1$ is the projective coordinate
on the $(\alpha+1)^{\rm st}$ $\IP^1$, \etc;
one sees directly that the intersection of the $\alpha^{\rm th}$
and $(\alpha+1)^{\rm st}$ spheres is the point 
$u_{\alpha}=u_{\alpha+1}=0$.

We would now like to use this data %$a_\alpha$, $\alpha=1,...,r$
specifying a resolution of the singularity to define a
gauged linear sigma model.
The fact that each $\IP^1$ of the resolution has the
structure of $\CO(-a_\alpha)$ leads us to
define the $r \times (r+2)$ charge matrix: 
\eqn\gencartan{
Q_{\alpha i } = -a_\alpha \delta_{\alpha i } + 
	\delta_{\alpha +1, i } + \delta_{\alpha-1, i}
}
where $0 \leq i \leq r+1$ and $1\leq \alpha \leq r$. 
We also denote by $C_{\alpha\beta}$ the $r\times r$ 
square matrix $-Q_{\alpha \beta}$ for 
$1\leq \beta \leq r$.  This is a symmetric matrix which 
we refer to as a {\it generalized Cartan matrix}.

In the remainder of this subsection we will gather some
mathematical facts about the singularity resolution
that will be of use to us in the sequel;
we also indicate the physical interpretation of some 
of these mathematical identities.

The ordered set of positive integers $a_\alpha$ 
appearing in the continued fraction expansion \contfrac\
can be used to generate two related sets of integers 
$q_i$ and $p_i$, $i=0,\dots, r+1$ 
via  the recursion relations: 
\eqn\ffr{\eqalign{
{p_{j-1} / p_j} &= [a_{j}, a_{j+1},\dots, a_r] \cr
{q_{j+1} / q_j} &= [a_j, \dots, a_1] 
\quad,\qquad 1 \leq j \leq r
}}
(where the fractions are in lowest terms). 
In addition we define $p_{r+1}=0$ and $q_0=0$. 
In particular \ffr\ for $j=1$ gives
$n=p_0$ and $p=p_1$.%
\foot{Using the recursion relations below one can show that  
$n=\det[C_{\alpha\beta}]$ 
while $p$ is the determinant of the first minor.}

The sequences of integers $p_j$, and $q_j$ will be 
very useful in what follows. The first notable 
fact is that,
for the minimal resolution having all $a_\alpha\ge 2$,
the vectors
\eqn\vgen{
  \coeff1n\vv_j = \coeff1n(q_j,p_j)\quad,\qquad
	j=0,...,r+1
}
are the $U(1)_\X \times U(1)_\Y $ $R$-charges of   a set of chiral operators
$\{\CT_{q_\alpha}\}$, $\alpha=1,...,r$, which generate
the chiral ring of the orbifold $\IC^2/\IZ_{n(p)}$
\refs{\HarveyWM,\MartinecTZ}.

To see this, note that the  vectors $\vv_i = (q_i, p_i)$ satisfy
a repackaged form of the recursion relation:
\eqn\recrela{
a_i \vv_i = 
	\vv_{i-1} + \vv_{i+1}  \qquad 1\leq i\leq r\ .
}
with the boundary conditions $q_0=0$, $q_1=1$ for the $q_i$,
and $p_{r+1}=0$, $p_r=1$ for the $p_i$. 
It thus follows that 
\eqn\expdbs{
\vv_{j} = q_{j} \vv_1 + B_{j} \vv_0 \ ;
}
here $B_\alpha$ are integers satisfying the recursion relation 
$a_j B_{j} = B_{j+1} + B_{j-1}$, $B_0=1$, $B_1=0$.  A solution 
of these recursion relations with the required initial 
conditions is $nB_j = p_j - p_1 q_j$, and hence 
\eqn\rach{
(q_j, p_j) = q_j \vv_1 + (0,nB_j) 
}
Now, for the minimal resolution 
when all the $a_\alpha>1$, 
the sequences $q_i$, $p_i$ satisfy 
\eqn\order{\eqalign{ 
	q_0=0 &< q_1=1 < \cdots < q_{r+1}=n \cr
	p_0=n &> p_1=p > \cdots > p_r=1 > p_{r+1}=0 \ ,
}}
and hence the vectors $\vv_i$ lie in the fundamental domain. 
However, from \Rchge\ we see that the R-charge of the 
chiral field in the $\kappa$ twisted sector is just 
\eqn\ktwist{
  {\kappa\over n} \vv_1~\mod~1\ ,
} 
and so we identify \rach\ with the R-charges of
the $\kappa=q_\alpha$ twisted sector.

Finally, to see that these are the {\it generators} of the 
chiral ring we proceed as follows. First note 
  that for all resolutions 
(minimal or not) it is true that
\eqn\alwaystrue{
  \frac{p_i}{q_i}>\frac{p_{i+1}}{q_{i+1}}\quad ,
} 
in other words, the slopes of successive R-charge vectors
is always decreasing.  This is because, by \recrela,
the $\alpha^{\rm th}$ vector lies between its neighbors.
Now consider the parallelogram spanned by $\vv_{i-1},\vv_{i+1}$
and draw the line through $\vv_{i-1}+\vv_{i+1}$. From this 
figure it is clear that if $a_i\geq 2$ then $\vv_{i-1},\vv_i,\vv_{i+1}$
form part of the   boundary of 
a convex region in the $(q,p)$ plane (known as
the {\it Newton boundary}). On the other hand, if $a_i=1$ 
$\vv_{i-1},\vv_i,\vv_{i+1}$ certainly do not form a convex boundary.
For the minimal resolution we can write all the R-charge 
vectors in the fundamental domain as positive integral combinations of the 
generating set of vectors $\vv_i$. Assuming the OPE 
coefficients are generically nonzero, we conclude that these 
twist fields are a set of generators of the chiral ring. 

Equation \recrela\ implies that one can write relations 
on the orbifold chiral ring,
\eqn\chiringrels{
  (\CT_{q_\alpha})^{a_\alpha}= \CT_{q_{\alpha-1}}\CT_{q_{\alpha+1}} 
	\quad,\qquad \alpha=1,\dots,r
}
(where we have defined $\CT_0=\CV_\Y$ and $\CT_{r+1}=\CV_\X$,
the untwisted sector ``volume form'' chiral operators
of the $X$, $Y$ planes of $\IC^2$).
These ring relations
simply encode the additivity of twist quantum numbers
in the $X$, $Y$ planes of $\IC^2$.

It follows from \recrela\ that 
$\vv_i \times \vv_{i+1} = 
\vv_{i -1} \times \vv_{i} $, and hence: 
\eqn\area{
q_i p_{i-1} - q_{i-1} p_i =n \quad,\qquad 1\leq i\leq r+1.
}
More generally, one can show that, for $i>j+1$, 
\eqn\gencross{
     q_i p_j - p_i q_j = n' n \quad,\qquad 0\leq i\leq r+1
}
where (for $i>j$) the continued fraction 
$[a_{j+1},...,a_{i-1}] = n'/p'$ determines $n'$.
We will make extensive use of these identities below.

There is also a nice formula for the inverse of the
generalized Cartan matrix $C_{\alpha\beta}$
in terms of the $(q_i,p_i)$:%
\foot{This equation
elegantly generalizes the standard formula for the inverse Cartan 
matrix of the $A_{r}$ Dynkin diagram defined by $a_\alpha=2$, where 
$q_i= i $ and $p_i = (r+1-i)$, $0\leq i \leq r+1$.} 
\eqn\invgencartan{
(C^{-1})_{\alpha\beta} 
	 = \cases{ {1\over n} q_\alpha p_\beta &
		$1\leq \alpha\leq \beta\leq r$ \cr
		& \cr
	{1\over n} p_\alpha q_\beta &
		$1\leq \beta \leq \alpha\leq r$ }\quad .
}
One easily proves this claim using \recrela\ and \area.%

Finally, we come to an important identity on continued fractions. 
Let us define  
$[x,y] = x- 1/y$ for any pair of 
{\it real} numbers $x$, $y$, and then define
multiple continued fractions via $[x,y,z]:= [x, [y,z]]$.%
\foot{Warning: The ordering of the brackets matters.} 
A simple computation shows that 
\eqn\addone{
[x+1, 1, y+1] = [x,y]\ ;
}
this is why the continued fraction expansion of $n/p$ is only unique
if all the $a_\alpha>1$.

Returning to the resolution of $\IC^2/\IZ_{n(p)}$, 
the minimal resolution of the
singularity is defined by the criterion that  all the $a_\alpha>1$.
As we have mentioned,  there are `non-minimal' resolutions of the singularity
obtained by blowing up the point of intersection of the 
$k^{\rm th}$ and $(k+1)^{\rm st}$ $\IP^1$'s
in the resolution chain.  Since the space was nonsingular
before this operation, one is blowing up a point on what is locally
$\IC^2$, and this results in a curve of self-intersection $-1$.
The effect on the continued fraction expansion is
\eqn\blowup{
  \frac np=[a_1,...,a_k,a_{k+1},...,a_r]        \longrightarrow
        \frac np=[a_1,...,a_k+1,1,a_{k+1}+1,...,a_r]\ .
} 
This expanded sequence may be used to define a charge matrix \gencartan\
and hence  a GLSM  
with $U(1)^{r+1}$ gauge group. 
One may readily check that the original sequences of integers
$p_\beta$, $q_\beta$, $\beta=1,...,r$ is unaltered, 
and that a new pair $p_*=p_k+p_{k+1}$
and $q_*=q_k+q_{k+1}$ is added. More precisely, the 
sequence of integers on the RHS of \blowup\ defines a 
set of vectors ${\bf \hat v}_I, I=0,\dots, r+2$ related to 
the original charge vectors by 
\eqn\newvectors{
\eqalign{
{\bf \hat v}_i & = {\bf  v}_i \qquad i=0,\dots, k \cr
{\bf \hat v}_{k+1} & = {\bf  v}_k +  {\bf  v}_{k+1}   \cr
{\bf \hat v}_i & = {\bf  v}_{i-1} \qquad i=k+2,\dots, r+2 \quad .}
}
%

%because by \newvectors\ the new vector $\hat\vv_{k+1}$
%lies between its neighbors $\hat\vv_{k}$, $\hat\vv_{k+2}$
%(so the slopes are indeed decreasing).
%
%In particular, when one blows up an extra curve,
%the extra FI parameter $\hat\zeta_{k+1}$ of the $U(1)^{r+1}$
%gauge theory
%controls the size of the extra $-1$ curve in the resolution;
%it is blown up when $\hat\zeta_{k+1}>0$.
Associated to $\hat \vv_{k+1}$ is a   dependent chiral field
$\CT_{\hat q_{k+1}}=\CT_{\hat q_k}\CT_{\hat q_{k+2}}$ in the chiral ring
of the orbifold twist fields.%
\foot{When it exists; it may happen that
the candidate operator $\CT_{\hat q_{k+1}}$ 
lies outside of unitarity bounds
on the chiral ring of the $\CN=2$ orbifold CFT.} 
The blowing up procedure can of course be repeated
any number of times.

%%%%%%%%%%%%%%%%%%%%%%%%%%%%%%%%%%%%%%%%%%%%%%%%%%%%%%%%%%%%%%%%%%%%%%%%%%

\newsec{Geometry of $U(1)^r$ quotients}

In this section we will discuss the geometry of the 
solution of the   D-term equations \Dterms\ for the 
charge matrix $Q_{\alpha i}$ defined in \gencartan. 
%
%
%can be slightly rewritten as
%
We will show that if
   $\zeta_\alpha<0$ for all $\alpha$ then this space is simply 
a $U(1)^r$ ``bundle'' over the orbifold $\IC^2/\IZ_{n(p)}$. 
(The quotation marks refer to the fact that the fibration 
degenerates over the origin, because $\IZ_n$ fixes the origin.) 
When some of the $\zeta_\alpha>0$ there is a topology change 
and we get a partial resolution of the singularity. 
If all $\zeta_\alpha>0$ then 
we have a $U(1)^r$ bundle over a toric resolution of
the singularity. When all the $a_{\alpha}>1$ this is 
the Hirzebruch-Jung, or minimal resolution of the 
orbifold singularity associated to the
continued fraction $n/p = [a_1, \dots , a_r]$.

One clear way to understand the geometry of the 
 quotient space  is to make a change of
basis on the generators of the $U(1)^r$ gauge group
so as to diagonalize the $U(1)^r$ action on the 
$X_\alpha$, $\alpha=1,...,r$.  One thus defines
\eqn\newphi{
\phi_{\beta} = C_{\beta\alpha} \theta_\alpha
}
so that the gauge rotation acts as
\eqn\gaugerot{
  X_\beta \to e^{i \phi_{\beta}} X_{\beta} 
}
with no sum, for $1\leq \beta\leq r$. 

This change of basis of the gauge group generators 
leads to a corresponding change in the D-term 
equations. We define 
\eqn\rmat{
%  R_{\alpha i} =  C^{-1}_{\alpha\beta} Q_{\beta i}
%	=-\frac{p_\alpha}n \delta_{i, 0}
%	-\frac{q_\alpha}n \delta_{i, r+1}
%	+\delta_{\alpha,i}\ \quad ,
  R_{\alpha i} = C^{-1}_{\alpha\beta} Q_{\beta i}
        = \frac{p_\alpha}n\delta_{i,0}
                +\frac{q_\alpha}n\delta_{i,r+1}
                        -\delta_{\alpha,i} \qquad,
}
where in the second equation we have used \invgencartan. 
Accordingly the D-term equations become 
\eqn\newdees{
 n R_{\beta i}|X_i|^2=
	p_\beta \vert X_0 \vert^2 - n \vert X_\beta \vert^2 
	+ q_\beta \vert X_{r+1}\vert^2 
		= n \zeta_\beta' \ .
}
where we define 
\eqn\lintrans{
  \zeta'_\alpha := C^{-1}_{\alpha\beta} \zeta_\beta\ .
}
Let us denote the   solution set  of \newdees\ by 
$\CS_{\vec \zeta}$. We are interested in the geometry 
of the quotient space $\CS_{\vec \zeta}/U(1)^r$. 

The different phases of the linear sigma model are again 
controlled by the signs of the $\zeta_\alpha$ 
({\it not} the $\zeta'_\alpha$, since as we shall see it is 
the former that are the physical sizes of the $\IP^1$'s
in the resolution chain). Nevertheless, as we discuss 
in the next section, discussions of the renormalization 
group flow are most naturally expressed in terms of $\zeta'_{\alpha}$
since their $\beta$ function 
 is directly related to the $R$-charges 
of the perturbations.

%%%%%%%%%%%%%%%%%%%%%%%%%%%%%%%%%%%%%%%%%%%%%%%%%%%%%%%%%%%%%%%

\subsec{Description of the quotient when all $\zeta_\alpha <0$}

Consider first the orbifold phase of large negative $\zeta_\alpha$.
 Since the matrix elements
of $C^{-1}$ are all positive, when all the $\zeta_\alpha$
are negative the  $\zeta'_\alpha$ are also negative. 
The D-term constraints then force $|X_\beta|>0$.
One may then fix the gauge 
analogous to \neighb\ by choosing $X_{\beta}>0$. 
That is, we can write the general element on the solution 
set in the form: 
\eqn\gfix{
%  (X_0,X_\beta,X_{r+1})=
%	\Bigl(e^{-i\sum_{\alpha} (p_\alpha/n)\phi_\alpha}\xi_1,
%	e^{i\phi_\beta}|X_\beta|,
%	e^{-i\sum_{\alpha} (q_\alpha/n)\phi_\alpha}\xi_2\Bigr)\ ,
  (X_0,X_\beta,X_{r+1})=
	\Bigl(e^{i \vec p\cdot \vec \phi/n}\xi_1,
	e^{-i\phi_\beta}|X_\beta|,
	e^{i\vec q\cdot\vec \phi/n}\xi_2\Bigr)
}
(where \eg\ $\vec p\cdot\vec\phi=\sum_\alpha p_\alpha \phi_\alpha$,
and $\beta=1,...,r$),
again up to a discrete quotient.  The unbroken discrete
gauge symmetry is in general a subgroup of $\IZ_n$ for a given $U(1)$,
acting as
\eqn\discgpact{
  (\xi_1,\xi_2) \sim (\omega^{p_{\beta}}\xi_1, \omega^{q_{\beta}}\xi_2)
}  
(it may happen that the greatest common divisor of 
$n$, $p_\beta$, and $q_\beta$ is greater than one).
Because of the identity \gencross,
the $(q_\beta)^{\rm th}$ power of the $\alpha^{\rm th}$ group action
\discgpact\ is identical to the $(q_\alpha)^{\rm th}$ power 
of the $\beta^{\rm th}$ group action \discgpact; 
the various group actions are simply
different elements of the {\it same} $\IZ_n$.
Our canonical choice is to consider the $\IZ_n$ to be generated
by \discgpact\ for $\beta=1$, with $p_1=p$ and $q_1=1$.
We conclude that if all $\zeta_\beta < 0$ then fixing the gauges 
$|X_\beta| >0 $ leaves unbroken a {\it single} $\IZ_{n(p)}$ gauge symmetry 
out of the $U(1)^r$ and hence the quotient is  $\IC^2/\IZ_{n(p)}$.

%%%%%%%%%%%%%%%%%%%%%%%%%%%%%%%%%%%%%%%%%%%%%%%%%%%%%%%%%%%%%

\subsec{Description in the region $\zeta_\alpha >0 $} 

It is trivial to solve the D-term equations in the form 
\newdees. Let us write the solution as 
\eqn\solmm{
  \vert X_\alpha\vert^2  = 
 	{p_\alpha\over n} \rho_1^2 + {q_{\alpha}\over n} \rho_2^2 
	- \zeta'_{\alpha} 
}
where again $\rho_i=|\xi_i|$, $i=1,2$.
In the region $\CD$ where 
\eqn\convexreg{
p_\alpha \vert X_0\vert^2 + q_{\alpha } \vert X_{r+1} \vert^2 
> n \zeta_{\alpha}'
%
%= n   C^{-1}_{\alpha\beta} \zeta_{\beta} 
%
}
for all $\alpha$, we can fix the gauge and parametrize the solution of the 
level set $\mu_\alpha= \zeta_\alpha$ by 
\eqn\paramtrze{
(X_0, X_1, \dots, X_{r+1}) = 
	(e^{i \theta_1} \xi_1, e^{-i \phi_1} \vert X_1\vert, 
	\dots,e^{-i \phi_r}  \vert X_r \vert , e^{i \theta_r} \xi_2 ) 
}
where $\phi_{\alpha} = C_{\alpha \beta} \theta_\beta$.

\bigskip
{\vbox{{\epsfxsize=4.5in
        \nobreak
    \centerline{\epsfbox{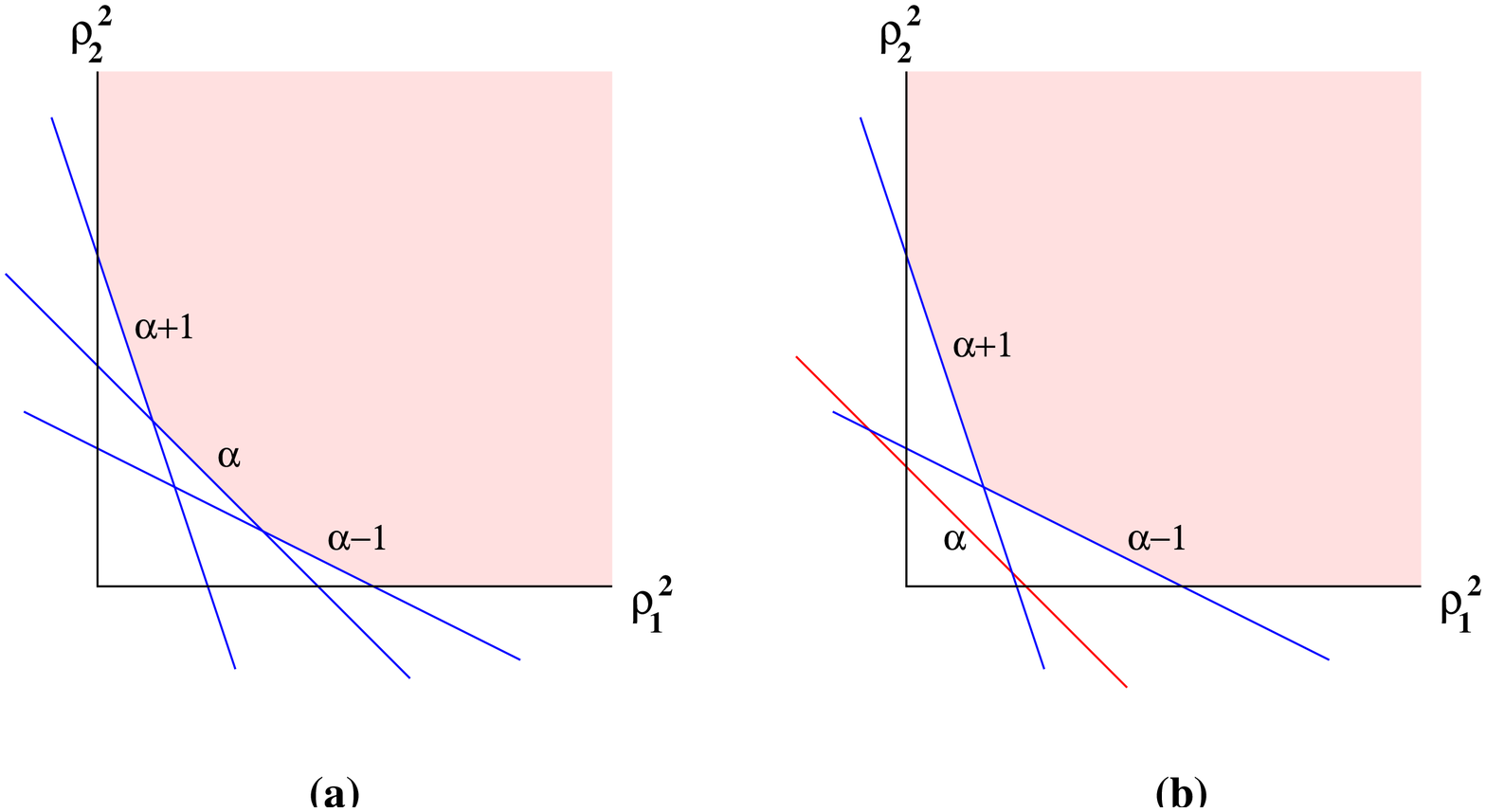}}
        \nobreak\bigskip
    {\raggedright\it \vbox{
{\bf Figure \rhobdy.}
{\it
(a) When all $\zeta_\alpha>0$ in the minimal Hirzebruch-Jung resolution, 
the convex region $\CD$ in the $\rho_1^2$-$\rho_2^2$ plane
allowed by the D-term constraints $|X_\alpha|>0$
is indicated by the shaded region.  
The $\alpha^{\rm th}$ segment is the longitudinal
direction of the minimal size curve $\CC_\alpha$
of the resolved space.
\hfil\break
(b) Blowing down, say, the curve $\CC_\alpha$ is achieved by adjusting the
associated FI parameter $\zeta_\alpha$ to negative values,
so that the constraint \convexreg\ is redundant; 
the line segment $|X_\alpha|=0$ shrinks away.
} }}}}
    \bigskip}

When we use the data of the minimal resolution, the region 
\convexreg\ describes a convex region $\CD$
in the $(\vert X_0\vert^2, \vert X_{r+1}\vert^2)$
plane which is indicated in figure \rhobdy.   
The region is convex due to the property \alwaystrue,
which says that the normals to the constraint boundaries have
monotonically decreasing slope.
The boundaries of $\CD$ are line segments where 
$\vert X_\alpha\vert=0$. On this line segment we cannot fix the gauge 
freedom as in \paramtrze. In particular, the $U(1)$ gauge freedom
associated with the angle $\phi_\alpha$ cannot be fixed. 

The line segment $\vert X_\alpha\vert=0$ is most usefully described 
as the equation 
\eqn\roundthree{
\vert X_{\alpha-1}\vert^2 + \vert X_{\alpha+1}\vert^2 = \zeta_{\alpha} 
}
with $0 \leq \vert X_{\alpha\pm 1} \vert^2 \leq \zeta_{\alpha}$.
On the interior of this segment all $\vert X_{\beta}\vert^2>0$ for 
$\beta\not=\alpha$, by convexity. Accordingly, we may partially fix the 
gauge by choosing $X_i>0$ for $i=0,\dots, \alpha-2$ 
and $i=\alpha+2,\dots, r+1$. 
Using \invgencartan\ and \gencross\ it is easy to show that 
$\phi_1,\dots, \phi_{\alpha-2}, \phi_{\alpha+2}, \dots, \phi_{r+1}$ are 
completely fixed, while the single remaining $U(1)$ gauge freedom is 
described by 
\eqn\unfixed{
\phi_{\alpha-1} = \phi_{\alpha+1} = -{1\over a_\alpha} \phi_{\alpha}
}
Therefore, the quotient of \roundthree\ by this remaining 
gauge freedom  is simply the standard Hopf fibration over $\IP^1$.
Thus, by including the boundary $X_{\alpha}=0$ into the K\"ahler 
quotient we are including a sphere, which we denote as $\CC_{\alpha}$. This 
sphere is the zero-section of the normal bundle with   $X_{\alpha}$ 
as holomorphic normal coordinate. From \unfixed\ we learn that the 
transition function of the normal bundle over $\CC_{\alpha}$ is 
$e^{-i a_\alpha \phi} $ where $\phi$ is the azimuthal coordinate 
on $\IP^1$, and hence the self-intersection number of $\CC_{\alpha}$ 
is $-a_{\alpha}$. The endpoints of the interval  $X_{\alpha\pm 1}=0$
describe the intersection with the spheres $\CC_{\alpha\pm 1}$. 
In this way one verifies again the fact that the intersection
form of the minimal resolution is 
\eqn\intersectionmtrx{
\CC_{\alpha}\cdot \CC_{\beta} = -C_{\alpha\beta}. 
}

The above description of $\CC_{\alpha}$ makes it straightforward to 
compute the volume of the 
curve $\CC_{\alpha}$ 
in the K\"ahler quotient metric. To do this we further
fix the gauge by requiring that $X_{\alpha-1}>0$. 
Setting $z= X_{\alpha+1}/X_{\alpha-1}$, 
a stereographic coordinate for $\IP^1$, we have 
\eqn\sovlea{
X_{\alpha-1} = + \sqrt{\zeta_{\alpha} \over 1+ \vert z\vert^2} 
}
Now, restricting the K\"ahler form 
\eqn\kahlrfrm{
\Omega = {i \over 2} \sum_{i=0}^{r+1} d X_i \wedge d \bar X_{i} 
}
to this gauge slice we get 
\eqn\restrict{
\iota^*(\Omega) = 
	\half d \bigl( \vert z\vert^2 X_{\alpha-1}^2\bigr) 
		\wedge d \theta
}
where $\theta$ is the phase of $z$. It follows that 
\eqn\volume{
\vol(\CC_{\alpha}) = \pi \zeta_{\alpha}.
}

One must use caution when interpreting \volume\ in the 
quantum theory. First of all the   kinetic terms in the sigma model limit
takes the metric of the target space away from
the metric induced by the K\"ahler quotient.  
Moreover, the K\"ahler quotient metric itself is only an 
approximation to the the renormalized spacetime in certain 
regions of spacetime. Indeed, the literal K\"ahler quotient 
metric is not ALE; rather, it has a fairly intricate 
structure at infinity. However, we should take the 
continuum limit $\Lambda\to + \infty, \zeta_{\rm bare} \to 
- \infty$, while working at finite values of the chiral 
fields $X_i$. In this regime we expect   the induced metric 
 to be a reliable {\it qualitative} guide to the geometry.

%%%%%%%%%%%%%%%%%%%%%%%%%%%%%%%%%%%%%%%%%%%%%%%%%%%%%%%%%%%%%%%%%%

\subsec{Homology, Cohomology, and K-theory for the toric resolutions}  

Let $\CX$ be a toric resolution of $\IC^2/\IZ_{n(p)}$ 
corresponding to $n/p=[a_1, \dots, a_r]$. We need not 
assume it is the minimal resolution. 

Since the homology and cohomology groups are 
homotopy invariants, we can compute them from 
the deformation retract of $\CX$ to the 
chain of spheres. It follows that:  
\eqn\torichom{\eqalign{
H_j (\CX )  & =  \cases{\IZ & $j=0$ \cr 
			\IZ^r & $j=2$ \cr
			0  & $j=4$ } \cr
 & \cr
H^j (\CX)  & =  \cases{\IZ & $j=0$ \cr 
			\IZ^r & $j=2$ \cr
			0  &  $j=4$ }\quad .
}}
As we have seen in the previous section, 
 the intersection form on $H_2$ is simply $-C_{\alpha\beta}$.

Now, Poincare duality for the smooth space $\CX$ says that 
$H^k(\CX) \cong H_{n-k}(\CX, \p \CX) \cong H_{n-k, cpt}(\CX) $ 
and $H^k_{cpt}(\CX) = H^k(\CX,\p \CX) \cong H_{n-k}(\CX)$. 
It follows that 
\eqn\toricccpthom{\eqalign{
H_{j,cpt} (\CX )  & =  \cases{0 & $j=0$ \cr 
			\IZ^r & $j=2$ \cr
			\IZ  & $j=4$ } \cr
 & \cr
H^j_{cpt} (\CX)  & =  \cases{0 & $j=0$ \cr 
			\IZ^r & $j=2$ \cr
			\IZ &  $j=4$ }\quad .
}}

Evidently, there is no torsion in the cohomology, and, by the 
Atiyah-Hirzebruch spectral sequence, no torsion in the K-theory.%
\foot{A nice description of the K-theory of 
toric varieties is given in \morelli. The author 
states that his results are only guaranteed for 
compact toric varieties. It would be nice to know 
if they apply to the noncompact case.} 
It follows that  
\eqn\OK{\eqalign{
K^0(\CX) &= \IZ \oplus \IZ^r \cr
K^0_{cpt}(\CX) &= \IZ \oplus \IZ^r \ .
}}
while $K^1=0$. 

There is a natural basis for $K^0(\CX)$ provided by the 
tautological line bundles. First, we take $\CO$, the trivial 
complex line bundle. This corresponds to a single $Dp$-brane 
wrapping the space $\CX$ with trivial Chan-Paton bundle and 
zero connection. Next, we construct tautological line bundles 
$\CR_{\alpha}$ corresponding to $Dp$-branes filling $\CX$
with magnetic monopoles in the different 
exceptional divisors $\CC_{\alpha}$. These are 
constructed as follows: 

In the geometrical phase we have a $G=U(1)^r$ principal 
bundle $\CS_{\zeta} \to \CX$. Let us denote 
a generic element $g \in G$ by 
\eqn\genelt{
  g= (e^{i \theta_1}, \dots, e^{i \theta_r}) := (g_1,\ldots,g_r)
} 
and acting on the chiral fields as 
\eqn\genXact{
  X_i\to \prod_\alpha g_\alpha^{Q_{\alpha i}}X_i \ .
}
We can define a natural 
collection of line bundles $\CR_{\alpha}$ as the   associated bundle to the
representation  $\rho_\alpha(g) = e^{i \theta_\alpha}$. These are
\eqn\canonline{
\CR_{\alpha} :=   \bigl( \CS_{\zeta} \times \IC \bigr)/U(1)^r
}
where the $\alpha^{th}$ $U(1)^r$ action is 
\eqn\uoneract{
g \cdot (X, v) = (X\cdot g, e^{-i \theta_{\alpha}} v) 
}

There is a canonical pairing of $K(\CX) \otimes K_{cpt}(\CX) \to \IZ$ 
given by the index theorem. Under the Chern isomorphism this is 
the same as the pairing $H^*(\CX) \otimes H^*_{cpt}(\CX) \to \IZ$. 
Thus, $\CO$ is dual to a D0-brane supported at a point on $\CX$, and there 
is a nondegenerate pairing between 4-branes with 2-brane charge 
$\CR_{\alpha}$ and 2-branes wrapping $\CC_{\alpha}$. 
These physical statements have precise mathematical analogs, and 
indeed a basis 
for the compactly supported derived category of $\CX$ (and hence 
of the compactly supported $K$-theory) is described in 
\refs{\wunram,\ishii,\riemenschneider}. 

Now, if we compare this with the equivariant K-theory appropriate 
to the orbifold then 
$$
K_\Gamma(\IC^2) = \IZ \oplus \IZ^{n-1}
$$
where the first summand corresponds to the regular representation 
and hence to a D0 brane, while the second summand corresponds to 
the fractional branes. Comparing this with the compactly 
supported $K$-theory of $\CX$ we see that for the non-supersymmetric 
case when $r< n-1$ we have a mismatch of the K-theory
of the orbifold and of its smooth resolution, and thus a general
statement of the problem of what happens to the extra $n-r-1$
topological charges, and the D-branes they couple to.
The resolution of this puzzle will be a generalization
of the one we found for the rank one case in section 3.
However, before we discuss it, let us first 
turn to a more detailed examination of the tautological bundles 
$\CR_{\alpha}$. 

%%%%%%%%%%%%%%%%%%%%%%%%%%%%%%%%%%%%%%%%%%%%%%%%%%%%%%%%%%%%%%%

\subsec{Canonical line bundles and special representations of $\IZ_n$ } 

In the mathematics literature \refs{\wunram,\ishii,\riemenschneider}
there is a statement of a generalized McKay correspondence under 
which the tautological bundles $\CR_{\alpha}$ on the {\it minimal 
resolution} correspond to so-called ``special representations'' 
of $\IZ_n$. In this section we will give a simple physical 
interpretation of this concept.

As we have seen,  
when all the $\zeta_{\alpha}>0$ there is a set of
tautological line bundles
associated with the unitary irreps $\rho_i$ of $G=U(1)^r$.  In the 
phase where $\zeta_{\alpha}<0$, the gauge group $U(1)^r$ is 
spontaneously broken to $\Gamma = \IZ_n $ by the nonzero VEV's 
of the chiral scalar fields $X_\alpha$.
We may identify this group $\Gamma$ with the orbifold 
group of the CFT. The representations $\rho$ of $G$
may be restricted to the subgroup $\Gamma$, 
and as such are unitary irreps of $\Gamma$. 
These are the ``special representations'' of the mathematical 
literature.%
\foot{A similar construction has been used by P. Mayr in 
\MayrAS.}
We will now explain this construction in more detail.

Consider the tautological bundles $\CR_{\alpha}$ in the 
geometrical phase, $\zeta_{\alpha}>0$ for all $\alpha=1,...,r$. 
If we ``continue'' in $\vec \zeta$ to the 
region with all $\zeta_{\alpha}<0$ 
what happens to the quotient space \canonline?% 
\foot{By ``continue'' we mean that one can consider the total space 
$\CS$ of the family of manifolds $\CS_{\zeta}$ 
over the base space of all $\zeta$. 
There is a $U(1)^r$ action on $\CS$ and we can form the associated 
line bundle over the quotient. We choose a single representation of 
$U(1)^r$ and compare the same line bundle restricted to
 the fiber over $\zeta>0$ and 
$\zeta<0$.}

As we have seen in sec. 5.1, 
in  the orbifold phase  the gauge group is broken to a
$\Gamma = \IZ_n $ subgroup of G.  Let us describe this
subgroup more precisely. The $X_{\beta}$  for $1\leq \beta \leq r$   have
nonzero vev's, so by \paramtrze\ one has
\eqn\unbrok{
\phi_{\beta} = 2\pi m_\beta
}
for integers $m_\beta $; these phases are related to
those of $U(1)^r$ in equation \genelt, via 
\eqn\restheta{
  \theta_{\alpha} = C^{-1}_{\alpha\beta} \phi_{\beta} = 
	{1\over n} \biggl( \sum_{\alpha< \beta} 
		q_{\alpha} p_\beta \phi_\beta + 
	\sum_{\alpha\geq \beta} p_{\alpha} q_\beta \phi_\beta \biggr)
}
Due to equations \gencross, \unbrok, one can replace 
$q_\alpha p_\beta$ by $p_\alpha q_\beta$ in the first sum,
modulo $2\pi\IZ$.  We then have that 
the unbroken subgroup is the set of elements
\eqn\unbrokensub{
  \Bigl(\exp[i \theta_1] , \dots, \exp[i \theta_r] \Bigr) =
	\Bigl( \exp[ 2\pi i  {p_1\over n} \vec q \cdot \vec m ],
	\exp[ 2\pi i  {p_2\over n} \vec q \cdot \vec m ] ,
		\dots,
	\exp[ 2\pi i  {p_r\over n} \vec q \cdot \vec m ] \Bigr)
}
as $\vec m $ ranges over all elements of $\IZ^r$.

Now, since $q_1 = 1$, the quantity $\vec q\cdot \vec m $
takes on all integral values.
Therefore, the unbroken subgroup is precisely the $\IZ_n$
subgroup described by integer powers of
\eqn\generator{
\hat g : =  \Bigl( \exp[ 2\pi i  {p_1\over n} ],
\exp[ 2\pi i  {p_2\over n}  ] ,
\dots,
\exp[ 2\pi i  {p_r\over n}   ] \Bigr)   \quad .
}

Let us now consider the ``evolution'' of the line bundle $\CR_{\alpha}$ 
associated to the gauge group $U(1)^r$ by the representation
\eqn\gprep{
  \rho_\alpha (g)  =   e^{i \theta_\alpha }
}
as we proceed form $\zeta_\alpha>0$ to $\zeta_\alpha<0$. 
    Evidently, this becomes a
line bundle associated to the unbroken gauge group
$\Gamma \subset U(1)^r$ (generated by \generator)
in the orbifold phase.
By equation \generator\  that representation is the  $p_\alpha^{th}$
power of ``the'' fundamental representation of $\IZ_n$.

The reason for the quotes 
is that there is some ambiguity in this statement 
since one must {\it choose} a generator of the dual group
$\widehat {\IZ}_n \cong \IZ_n $  in order to speak of
the $p_{\alpha}^{th}$ power of a generator of
$\widehat{\IZ}_n$. Since $p_r = 1$, we  are implicitly
choosing the generator $\rho_f$ that takes  
\eqn\ghat{
  \hat g \to exp[2\pi i /n]\ ,
}
and then the special representations are
\eqn\rhosy{
  \rho_\alpha =  (\rho_f)^{p_\alpha}
}
which is the main result of this subsection.%
\foot{One could however make other choices.  For example,
one could consider the resolution chain in the reverse order,
via $n/p'=[a_r,a_{r-1},...,a_1]$.  Geometrically
this is the same space, and one readily sees that
$p'=q_r$ of the original sequence $[a_1,...,a_r]$.
By \gencross, $p'p=p_rq_1=1$ mod $n$, and so this choice of
generator of $\IZ_{n(p)}$ is simply the $p'$ power of $\hat g$.
The result of reversing the sequence of $a_\alpha$
is to interchange the $p$'s and $q$'s, so that
the $\IZ_n$ representations associated to the
line bundles become $\rho_f^{q_\alpha}$.}

In summary, 
the natural line bundles $\CR_{\alpha}$ of the 
resolved Hirzebruch-Jung space ``analytically continue'' 
to the $r$ ``special representations'' of $\IZ_n$ in the quiver picture
of the orbifold.  We think this gives 
a nice picture of the ``generalized McKay correspondence,'' 
advocated in \refs{\wunram,\ishii,\riemenschneider}.
Having understood this, we also see that  
 the ``special'' representations just defined 
are not {\it so} special.  By additional blowups
of the sort described at the end of section 4,
one introduces additional R-charge vectors $\vv_*=(q_*,p_*)$
associated to the resolution of the singularity.
If the $p_*$ do not belong to the set $\{p_1,...,p_r\}$
of the minimal resolution, we can account for 
more of the $\IZ_n$ representations associated to fractional
branes of the orbifold.  In fact, this is only part of
the story, as we will see shortly.

%%%%%%%%%%%%%%%%%%%%%%%%%%%%%%%%%%%%%%%%%%%%%%%%%%%%%%%%%%%%%%%

\subsec{Connections on the canonical line bundles $\CR_{\alpha}$ } 

In order to write boundary interactions in the GLSM 
we need an actual connection 
on the line bundles $\CR_{\alpha}$.  In this section we will write formulae 
for one natural set of connections, $\Theta_{\alpha}$. We hope 
these will prove useful in future studies of the fate of D-branes 
in the geometrical phase using the methods of 
\refs{\HoriCK,\HoriIC,\KachruAN,\GovindarajanEF,\HellermanCT,
\GovindarajanKR,\DistlerYM,\KatzGH}.

A general principle gives us a natural $G$-connection on the 
principal $G$ bundle $\CS_{\zeta}$. 
Whenever there is a    $G$-invariant metric 
on a principal $G$-bundle $P$, there is a natural connection: The 
horizontal subspaces are the orthogonal complements to the 
$G$-orbits.  In terms of connection 1-forms the canonical 
vector fields define a map $\CB: \lieg \to TP$. The metrics 
allow us to define $\CB^\dagger: TP \to \lieg$, that is, a 
Lie-algebra-valued 1-form. The connection form is 
\eqn\connform{
\Theta = {1\over \CB^\dagger \CB} \CB^\dagger
}
Applying this principle to the present example
we get the following: 
Define 
\eqn\upstairsconn{\eqalign{
\tilde \Theta_{\alpha} & = 
	{i\over 2} \sum_{i} Q_{\alpha i} 
		(\bar X_i d X_i  - X_i d \bar X_i) \cr
	\CN_{\alpha  \beta} &= 
		\sum_i Q_{\alpha i} Q_{\beta i} \vert X_i\vert^2\ ;
}}
then 
\eqn\HJconnform{
  \Theta_{\alpha} =  \CN^{-1}_{\alpha \beta} \tilde\Theta_{\beta}
}
is the natural connection on the  line bundle $\CR_{\alpha}$.%
\foot{Note that $\CN_{\alpha \beta}$ is just the mass matrix for 
the $\sigma_{\alpha}$ fields in \Xpotl.}

The connection \HJconnform\ arises naturally as
a boundary interaction in the gauged linear sigma model 
\refs{\HoriCK,\HoriIC}.
The generalization of the treatment of \HoriCK\ section 6
%HIV eq 6.31 
to multiple gauge fields is
\eqn\bdyLone{
   L_{bdy} = \frac i2 \frac{n_\alpha}{\zeta_\alpha}
	Q_{\alpha i} (\bar X_i D X_i - X_i D \bar X_i)
                        - n_\alpha v_\alpha \ .
}
Here $v_\alpha$ are the $U(1)^r$ gauge fields
(the supersymmetrization of this interaction is
discussed for instance in \HoriIC).
The covariant derivative 
$D X_i = \partial X_i + Q_{\alpha i} v_\alpha X_i$ leads to
\eqn\bdyLtwo{
L_{bdy} = \frac i2 \frac{n_\alpha}{\zeta_\alpha} 
	Q_{\alpha i} (\bar X_i \partial X_i-X_i\partial \bar X_i)
        + \half v_\beta \Bigl[\frac{n_\alpha}{\zeta_\alpha} 
	\CN_{\alpha\beta} - n_\beta\Bigr]\ ,
}
and the constraint imposed by $v_\alpha$ on the boundary gives
$n_\alpha/\zeta_\alpha = \CN^{-1}_{\alpha\beta} n_\beta$.
Plugging back into $L_{bdy}$ gives
\eqn\bdyLthree{
L_{bdy} = \frac i2 n_\alpha (\CN^{-1})_{\alpha\beta}
           Q_{\beta i} (\bar X_i \partial X_i - X_i \partial \bar X_i)
} 
which is the gauge field \HJconnform.  
The integers $n_\alpha$ are the induced $D(p-2)$ charges 
carried by a $Dp$ brane parallel to the Hirzebruch-Jung space.

The curvatures $\pi^*(F_\alpha)=d\Theta_{\alpha}$ 
generate $H^2(\CX)$, and moreover are
dual to the the two-cycles $\CC_{\alpha}$ of the
resolved space:
\eqn\Fintsi{
  \int_{\CC_\beta} {F_\alpha \over 2\pi} = \delta_{\alpha\beta} \ .}
Let us pause to demonstrate this result.
Describe the curve $\CC_\beta$
by the region $X_{\beta}=0$ where 
we fix the gauge (away from the poles) 
so that all other $X$'s are positive, except for 
$X_{\beta+1} = z X_{\beta-1}$. 
It is then straightforward to see that 
\eqn\sovlebemin{
X_{\beta-1} =\sqrt{ {\zeta_\beta \over 1+ \vert z\vert^2} }\ ,
}
and all the other $X_\gamma$ may be solved for in terms of $z$.
Now we integrate the $(1,1)$-form
$d\Theta_{\alpha}$ over this region in $X$-space; 
after some algebra one finds
\eqn\Fintermediate{
  \int_{\CC_{\beta}}F_{\alpha} = \int_{\IC} 
	{\p \over \p \rho^2} \biggl( \rho^2 X_{\beta-1}^2 
	(\CN^{-1})_{\alpha,\gamma}Q_{\gamma, \beta+1}
	\biggr) d(\rho^2) \wedge d \theta 
}
where $z = \rho e^{i \theta} $.
Evaluating the total derivative, 
there is no contribution at $\rho^2=0$, 
and the contribution at $\rho^2=\infty$ is just 
\eqn\finalint{
  2\pi \zeta_{\beta} \Bigl.\Bigl( 
	(\CN^{-1})_{\alpha,\gamma}
	Q_{\gamma, \beta+1}\Bigr)
		\Bigr\vert_{X_\beta=0\cap X_{\beta-1}=0} \ .
}
In appendix A we show that 
\eqn\niceident{
  (\CN^{-1})_{\alpha\gamma} Q_{\gamma,\beta+1}
        = |X_{\beta+1}|^{-2} \delta_{\alpha\beta} 
	={1\over \zeta_{\beta}} \delta_{\alpha,\beta}
}
at the point $X_\beta=X_{\beta-1}=0$, and so
indeed $\int_{\CC_\beta} F_\alpha=2\pi \delta_{\alpha\beta}$
as claimed.

Since the boundary interaction \bdyLone\ has an 
$\CN=2$ supersymmetric completion, it follows 
that the curvatures $F_\alpha$ are type $(1,1)$. 
We do not know if they are Hermitian-Yang-Mills. 
If we introduce a boundary into the GLSM then there is 
simultaneous boundary RG flow along with the bulk RG 
flow of the localized tachyons. We expect that $F_{\alpha}$ 
will become non-normalizable for those curves that 
expand out to infinite radius while the $F_{\alpha}$ 
corresponding to the $-2$ curves will flow to the 
Hermitian-Yang-Mills connections of Kronheimer and Nakajima
\kronheimer. 

The reader should be warned that the naive D-brane charge 
formula need not apply in this case since the closed string 
background is not on-shell. Indeed, the integral 
\eqn\pairing{
\int_{X} {F_{\alpha}\over 2\pi} \wedge {F_{\beta}\over 2\pi} 
}
may be carried out explicitly, and is {\it not} $-C^{-1}_{\alpha\beta}$. 
There is a nontrivial contribution from the Chern-Simons term at 
infinity. It would be very interesting to know if there is a 
generalization of the standard Chern-Simons coupling to D-branes 
appropriate to the examples we are discussing. 

%%%%%%%%%%%%%%%%%%%%%%%%%%%%%%%%%%%%%%%%%%%%%%%%%%%%%%%%%%%%%%%%%%%
%%%%%%%%%%%%%%%%%%%%%%%%%%%%%%%%%%%%%%%%%%%%%%%%%%%%%%%%%%%%%%%%%%%

\newsec{RG flow of the Higgs branch }

The analysis of the previous section explains precisely
which RR charges associated to fractional D-branes are
``lost'' when the orbifold singularity is resolved by
tachyon condensation to its Hirzebruch-Jung minimal resolution.
The Hirzebruch-Jung space is the Higgs branch 
of the GLSM configuration space.
In the next section we will show that these ``missing''
RR charges are recovered by a careful analysis of the 
Coulomb branch of the GLSM, generalizing the rank one 
case analyzed in
section 3.  Actually, it will turn out that there is
an interesting interplay between the Higgs and Coulomb
branches of the GLSM that takes place under RG flow,
where topological charge can pass from the Higgs branch
to the Coulomb branch along the flow to the IR. 
We begin the story with a description of the RG flows
of the Higgs branch.

Consider a generic (\ie\ not necessarily minimal)
resolution of $\IC^2/\IZ_{n(p)}$ given by the GLSM
of the previous section.  The geometry flows with RG scale
according to the flow of the FI parameters $\zeta_\alpha(\mu)
= Re(t_{\alpha,{\rm eff}}(\mu))$ given in \FIren. 
To say that a certain collection of curves is blown up
one should specify the worldsheet scale at which 
the target space geometry is being considered.  
Individual curves of the resolution may blow up or down along the flow,
or remain of fixed size.

It turns out that the discussion of the renormalization 
group flow is much clearer once one diagonalizes the 
$U(1)^r$ gauge group action as in the previous section (see eq. \gaugerot)
and accordingly uses the FI parameters   $\zeta'_\alpha$. These 
satisfy the RG flow  
\eqn\primerun{\eqalign{
  \zeta^{\prime}_{\alpha,\eff} (\mu) &= 
	\zeta'_{\alpha,{\rm bare}} 
	-(1-\Delta_\alpha)\log\bigl({\mu\over \Lambda}\bigr)\cr
%
%\log\Bigl((1-\Delta_\alpha)|\sigma_\alpha|/\mu\Bigr)\cr
%
  \Delta_\alpha &= \frac{p_\alpha}n+\frac{q_\alpha}n\ .
}}
We recognize that $1-\Delta_\alpha$ are the scale dimensions
of the couplings $\lambda_\alpha$ 
to the twist operators $\CT_{q_\alpha}$
in the chiral ring of the orbifold CFT \znptwist. 
As discussed in section 4,   $\coeff1n(q_\alpha,p_\alpha)$ are
 the R-charges
of this operator.

A convenient way to picture the resolution is given in 
figure \rhobdy a.  The D-term constraints imply the inequalities
\convexreg, 
whose intersection is a convex region $\CD$
in the $\rho_1^2$-$\rho_2^2$ plane (recall that $\rho_{1,2}$
are the magnitudes of the gauge invariant coordinates $\xi_{1,2}$
on the target manifold).  The boundary $\partial\CD$ has two
semi-infinite segments corresponding to $\rho_{1,2}=0$,
and finite line segments corresponding to the curves
blown up in the resolution.  Roughly speaking, each
curve is a circle bundle (of the azimuthal direction)
over the corresponding line segment (the longitudinal direction).
The finite boundary segments of $\partial\CD$ are along
the straight lines $|X_\alpha|^2=0$ 
\eqn\dbdy{
  \frac{p_\alpha}n\rho_1^2+\frac{q_\alpha}n\rho_2^2=\zeta'_\alpha(\mu)
}
in the $\rho_1^2$-$\rho_2^2$ plane, 
where the running FI parameter is given by \primerun. 
%
%\eqn\zetamu{
%  \zeta'_\alpha(\mu)=
%	\zeta'_\alpha(\Lambda)-(1-\Delta_\alpha)\log(\mu/\Lambda)\ .
%}
%
The $\alpha^{\rm th}$ boundary thus moves with a speed 
$\frac{\partial}{\partial \log\mu}\zeta'_\alpha(\mu)$
along the flow, in the direction $(p_\alpha,q_\alpha)$.  

To see whether a given segment of the 
boundary corresponding to the $\alpha^{\rm th}$ $\IP^1$
is shrinking or growing along the flow, one needs to
determine whether the endpoints of that segment are
moving toward or away from one another.
The intersection of the $\alpha^{\rm th}$ and the
$\beta^{\rm th}$ boundaries is at the point
\eqn\bdyint{
	\CP_{\alpha\beta}:=
  \left.(\rho_1^2,\rho_2^2)\right|_{X_\alpha=X_\beta=0} =
	\frac{n}{p_\alpha q_\beta-q_\alpha p_\beta}
		\bigl(q_\beta\zeta'_\alpha(\mu)-q_\alpha\zeta'_\beta(\mu),
		p_\alpha\zeta'_\beta(\mu)-p_\beta\zeta'_\alpha(\mu)\bigl)\ .
}
The velocity of this point along the RG flow is
\eqn\intvel{\eqalign{
  \uu_{\alpha\beta}&=\frac{\partial}{\partial \log\mu}\CP_{\alpha\beta}\cr
	&= \frac{n}{q_\alpha p_\beta - p_\alpha q_\beta}
	\Bigl(q_\beta(1-\Delta_\alpha)-q_\alpha(1-\Delta_\beta)\;,\;
	p_\alpha(1-\Delta_\beta)-p_\beta(1-\Delta_\alpha)\Bigr)\ .
}}
In particular, 
\eqn\speccse{
\uu_{\alpha,\alpha-1} = (q_{\alpha-1} - q_{\alpha} + 1, 
p_{\alpha} - p_{\alpha-1} + 1) \ .
}
One way to determine whether 
the two endpoints of an interval are separating
or approaching is to consider the relative slopes of their
corresponding velocity vectors.  One finds that the
condition that two endpoints $\CP_{\alpha-1,\alpha}$
and $\CP_{\alpha,\alpha+1}$ are separating,
so that $\CC_\alpha$ is growing in size along
the RG flow, is that
%\eqn\sepcond{\eqalign{
%  0 &< \vv_{\alpha+1}\times\vv_{\alpha-1}-\vv_\alpha\times\vv_{\alpha-1}
%	-\vv_{\alpha+1}\times\vv_\alpha  \cr
%	&\qquad\qquad+n(2p_\alpha-p_{\alpha-1}-p_{\alpha+1})
%	+n(2q_\alpha-q_{\alpha-1}-q_{\alpha+1})\cr
%  &=n(a_\alpha-2)(1-\Delta_\alpha)\ ,
%}}
\eqn\sepcond{
\uu_{\alpha,\alpha-1}\times \uu_{\alpha+1,\alpha} = 
  n(a_\alpha-2)(1-\Delta_\alpha)>0\ ,
}
where we have used \area\ and \gencross. 

We see from this result that the curves of the minimal
resolution, once blown up, stay blown up;
\sepcond\ implies that in the
minimal resolution, $-2$ curves remain of fixed size
(their boundary segments remain of fixed length),
and curves with self-intersection $-3$ and below  grow in size.  
This is illustrated in figure \genericflow\
for the example of $n(p)=10(3)$, whose minimal resolution
corresponds to the continued fraction $10/3=[4,2,2]$.
The general picture of the RG flow can thus be summarized
as a splitting of the continued fraction expansion
\eqn\splitup{
  [\,\cdots, \underbrace{2,\ldots,2}_{\ell_1},\cdots,
	\underbrace{2,\ldots,2}_{\ell_2} ,\cdots\,]
  \longrightarrow[\,\underbrace{2,\ldots,2}_{\ell_1}\,]\oplus
        [\,\underbrace{2,\ldots,2}_{\ell_2}\,]\oplus\cdots
}
where all $a_\alpha>2$ in between the subsequences
of $a_\alpha=2$.  Thus the IR limit is a collection
of ALE spaces $A_{\ell_1}\oplus A_{\ell_2}\oplus\cdots$.

\bigskip
{\vbox{{\epsfxsize=5in
        \nobreak
    \centerline{\epsfbox{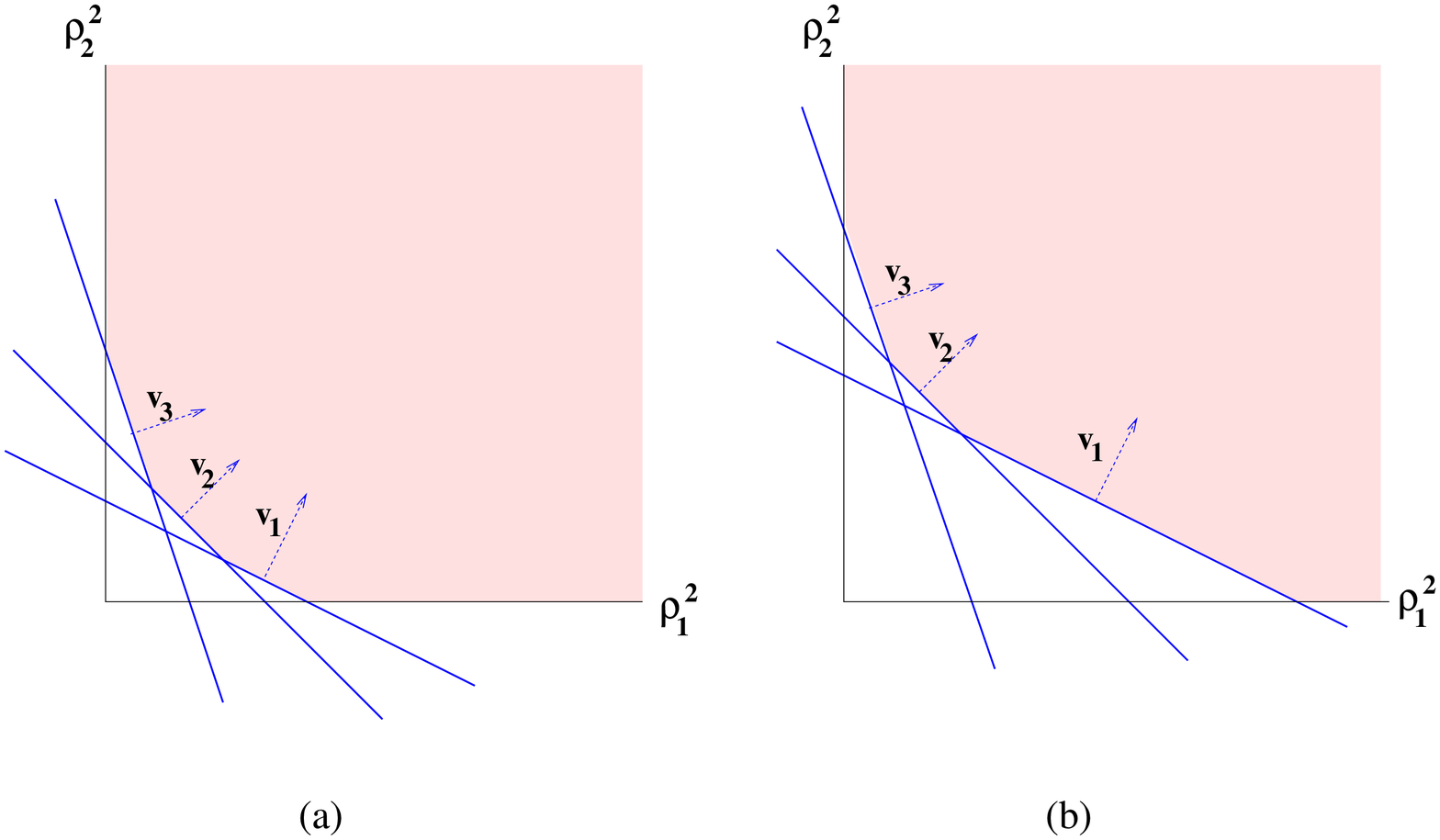}}
        \nobreak\bigskip
    {\raggedright\it \vbox{
{\bf Figure \genericflow.}
{\it
Generic RG trajectory for $n(p)=10(3)$, for which $10/3=[4,2,2]$
specifies the minimal resolution.  The $-4$ curve blows
up to infinite size along the flow, while the two $-2$ curves
remain of fixed size.
} }}}}
    \bigskip}

Of course, if we send an FI parameter
$\zeta'\to-\infty$, 
its associated wall moves down and to the left toward 
infinity and stays there under RG flow,
and imposes no constraint.
For instance, if we turn on only a single relevant perturbation
$\Sigma'_\alpha$ starting from the orbifold fixed point
(it need not even be one associated to a curve in
the minimal resolution), RG flow will make a single
curve that grows in size.  The convex region $\CD$
will be bounded by the walls $\rho_1=0$, $\rho_2=0$,
and the curve $|X_\alpha|=0$, see figure \specialflow.  
The endpoints of
the $|X_\alpha|=0$ segment will move up the $\rho_{1,2}$ axes
along the flow to the IR.  Since the remaining curves
of the resolution remain blown down, there will be
``daughter'' singularities at the north and south poles of $\CC_\alpha$.
This gives a picture of
the formation and separation of such daughter
singularities, previously analyzed in
\refs{\AdamsSV,\HarveyWM,\VafaRA,\MartinecTZ}.

\bigskip
{\vbox{{\epsfxsize=2.5in
        \nobreak
    \centerline{\epsfbox{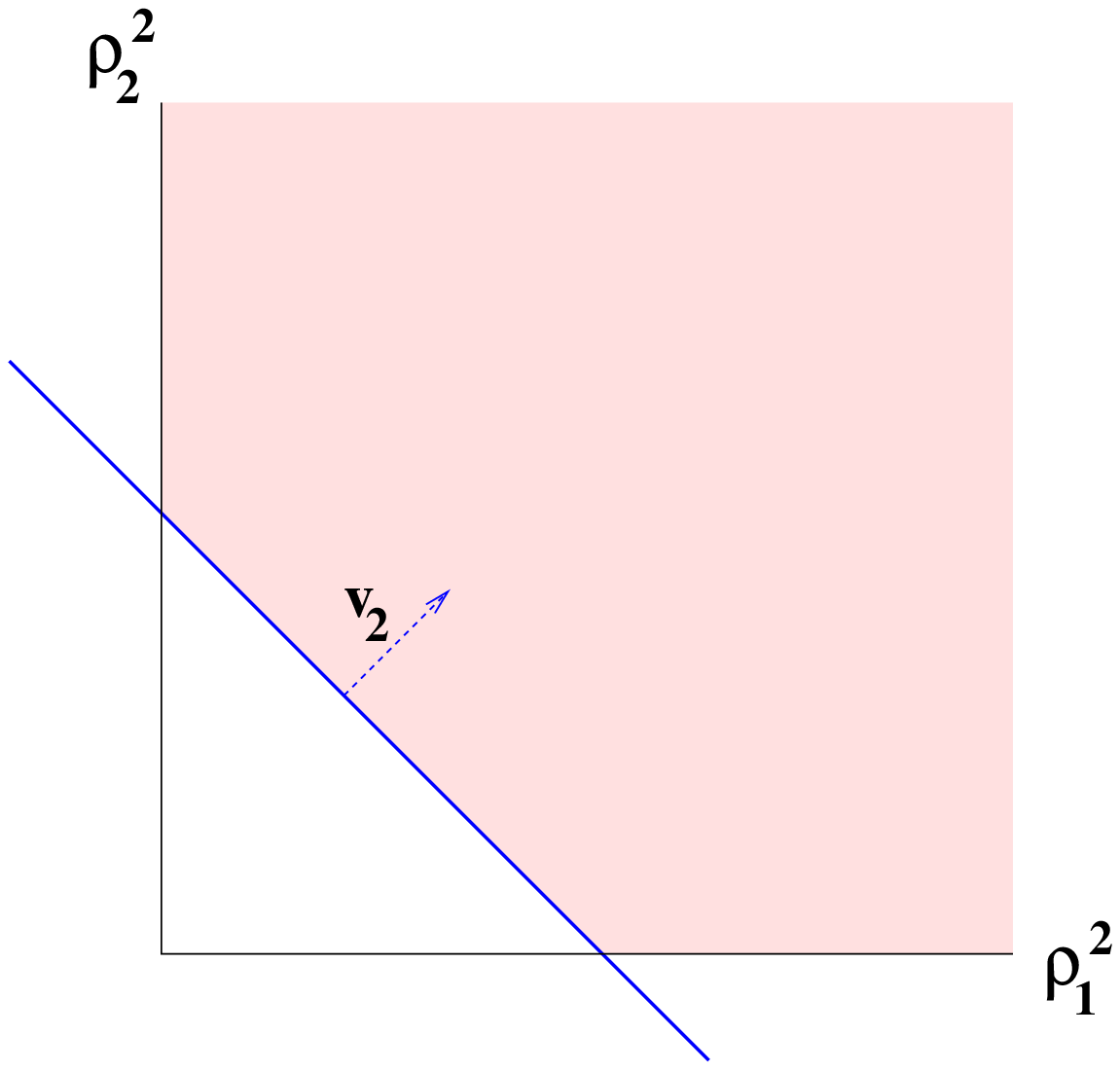}}
        \nobreak\bigskip
    {\raggedright\it \vbox{
{\bf Figure \specialflow.}
{\it
The special RG flow for the minimal resolution
of $n(p)=10(3)$, in which only the coupling $\zeta'_2$
is allowed to flow out of the orbifold point.
Two daughter singularities, $n'(p')=4(1)$ and $n''(p'')=2(1)$,
sit at the poles of $\CC_2$ and separate to infinite distance
along the flow.
} }}}}
    \bigskip}

%%%%%%%%%%%%%%%%%%%%%%%%%%%%%%%%%%%%%%%%%%%%%%%%%%%%%%%%%%%%%%%%%%%

\subsec{Non-minimal resolutions of the singularity: 
RG flow prunes the Higgs branch}

In sections 4 and 5.4, 
we discussed the effect of an additional blowup
of the resolved Hirzebruch-Jung singularity
\eqn\blowup{
  \frac np=[a_1,...,a_k,a_{k+1},...,a_r]	\longrightarrow
	\frac np = [a_1,...,a_k+1,1,a_{k+1}+1,...,a_r]\ .
}
The space with the extra blowup can be realized
as a $U(1)^{r+1}$ GLSM with charge matrix \gencartan\
determined by the blown up sequence.
The extra R-charge vector $\vv_*$ is
$\vv_*=(q_*,p_*)=\vv_k+\vv_{k+1}$.
There is an extra canonical line bundle associated to
the additional $U(1)$ gauge group in the GLSM
via the construction of section 5, which continues
from the geometrical phase to the orbifold phase
as the line bundle associated to the $\IZ_n$ representation
$(\rho_f)^{p_*}$.  Thus, by additional blowups one might
think that we can keep track
of more than the `special' $\IZ_{n}$ representations
of the minimal resolution of the singularity.
Alas, the condition \sepcond\ shows that all such
additional curves are blown down under RG flow to the IR,
see figure \rhoflow.

\bigskip
{\vbox{{\epsfxsize=4.5in
        \nobreak
    \centerline{\epsfbox{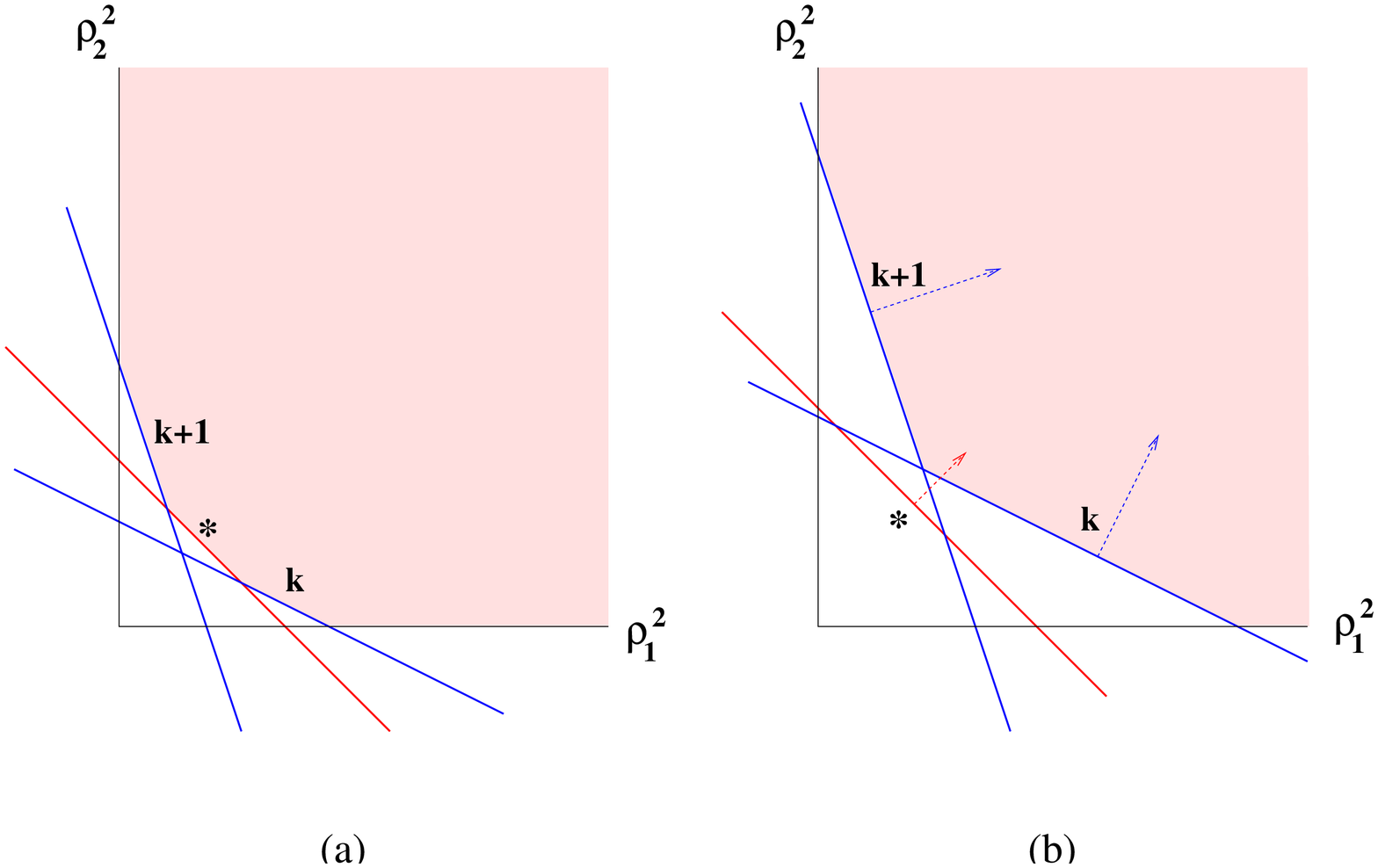}}
        \nobreak\bigskip
    {\raggedright\it \vbox{
{\bf Figure \rhoflow.}
{\it
(a) By fine tuning the FI parameters, in some range of 
worldsheet scales one can arrange that an extra curve
is blown up in the singularity resolution.
\hfil\break
(b) RG flow blows down the extra curve.
} }}}}
    \bigskip}

Equation \sepcond\ says that if an  $a_\alpha=1$
curve corresponds
to a relevant operator $1-\Delta_\alpha>0$, then
the line segment that defines
the minimal size curve $\CC_\alpha$ shrinks away,
because the flow of the neighboring walls outcompetes the
flow of the $\alpha^{\rm th}$ wall \dbdy;
and if the corresponding operator is irrelevant, 
the boundary in figure \rhoflow\ is moving to the lower left
and disappears from the geometrical region $\rho_i^2>0$
altogether.
One can also easily check that for multiple blowups, the
flow blows down in succession all the additional curves 
beyond those of the minimal resolution,
until in the far IR of the RG flow one is left
with the minimal resolution.%
\foot{One might have worried that blowing up twice,
\eg\ $[...,a_k,a_{k+1},...]\to[...,a_k+2,1,2,a_{k+1}+1,...]$,
turns a $-1$ curve into a $-2$ curve which is then stable
under RG flow.  However, the movement of the walls 
in figure \rhoflow\ is controlled by the R-charge vectors, 
which haven't changed.  One has added an extra segment
to the boundary in figure \rhoflow a, but the walls formed
by the $k^{\rm th}$ and $(k+1)^{\rm st}$
D-term constraints are still closing in.}

Thus, one may choose an RG trajectory for which,
in some range of worldsheet scales $\mu$, 
additional curves beyond the minimal
resolution are blown up.  In that range of scales, these
curves are part of the resolved geometry $\CX$ and therefore
of the Higgs branch of the GLSM configuration space.
Eventually, however, these curves are blown down in 
a generic RG flow; in a sense, the scaling operators $\CT_*$
which couple to the extra curves 
are `not as relevant' as those which blow up
the curves of the minimal resolution, and thus can't compete
with them in the long run.  

The precise sense in which these operators are
`not as relevant' is shown in figure \newton.
As discussed in section 4,
the generators of the chiral ring 
are in one-to-one correspondence with the curves
of the minimal resolution, and their R-charges
define the polygonal boundary (the Newton boundary)
of a convex region in the space of R-charges.
In other words, the line passing through any
pair of points $\vv_\alpha$, $\vv_{\alpha+1}$
has the R-charges of all other operators lying above it.  
In this sense, the operators corresponding to the
curves of the minimal resolution are the `most relevant'.

\bigskip
{\vbox{{\epsfxsize=2.1in
        \nobreak
    \centerline{\epsfbox{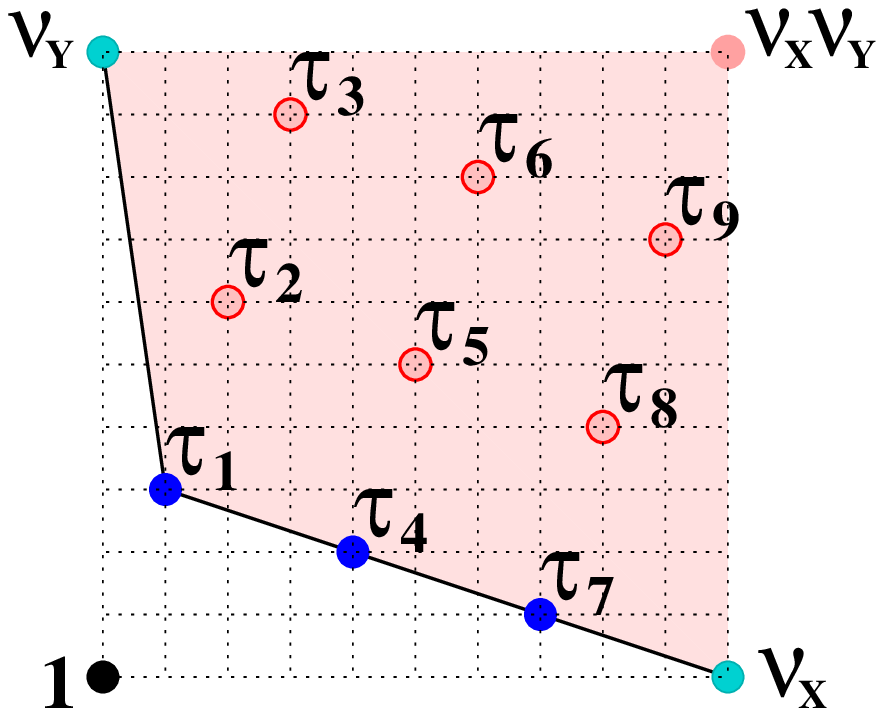}}
        \nobreak\bigskip
    {\raggedright\it \vbox{
{\bf Figure \newton.}
{\it
The Newton boundary of chiral operators
for $n(p)=10(3)$,
is the solid line bounding the shaded polygon.
The generators of the chiral ring 
lie on the Newton boundary, and all other chiral operators
lie above it.
} }}}}
    \bigskip}

%%%%%%%%%%%%%%%%%%%%%%%%%%%%%%%%%%%%%%%%%%%%%%%%%%%%%%%%%%%%%%%%%%%
%%%%%%%%%%%%%%%%%%%%%%%%%%%%%%%%%%%%%%%%%%%%%%%%%%%%%%%%%%%%%%%%%%%

\newsec{The Coulomb branch of the GLSM }

The results of the previous section show that  
 all line bundles but those associated to the ``special''
representations of $\IZ_n$ disappear from the purview
of the Higgs branch of the GLSM under RG flow.  However, the associated
topological charges do not simply disappear from the 
full GLSM configuration space.
%The general resolution of our paradox 
%lies in carefully keeping track
%of the worldsheet supersymmetric (RR) ground states of the GLSM.
Each RR gauge field of the orbifold CFT
is built from a vertex operator that creates
one of the supersymmetric ground states of the orbifold;
if we can follow the ground states, we can follow
the topological charge.  

The simplest way to characterize the 
RR ground states not present in the Higgs branch
is through the effective twisted chiral superpotential \renspotl.  
To analyze them, it is convenient to introduce (following \HoriKT)
a set of twisted chiral fields $Y_i$ which dualize 
(\`a la  Buscher \refs{\BuscherQJ,\RocekPS}) 
the phases of the $X_i$.
The basic feature of the duality transformation that we will use
is its effect on the twisted superpotential
\eqn\horivafa{\eqalign{
        \widetilde W&= \sum_{\alpha=1}^{r} \Sigma_\alpha
\Biggl(\sum_{i=0}^{r+1} Q_{\alpha i} nY_i-t_\alpha(\mu)\Biggr)
                        +\mu \sum_i \lambda_i e^{-nY_i}\cr}
}
where
%
%\eqn\lamdef{
%  \lambda_{\alpha}=\exp[C^{-1}_{\alpha\beta}t_\beta]
%        \equiv exp[\zeta'_\alpha]
%}
%and $\lambda_{0},\lambda_{r+1}\equiv 1$.
%
\eqn\teeflow{
t_{\alpha}(\mu):= t_{\alpha,{\rm bare}} 
+ \sum_{i=0}^{r+1} Q_{\alpha i} \log\bigl({\mu\over \Lambda}\bigr) \ .
}

Eliminating the $Y_i$ by their
equation of motion gives back \renspotl.
Instead, we will eliminate the $\Sigma_\alpha$ and $Y_\alpha$, 
$\alpha=1,...,r$, by their equations of motion to get
(in terms of $u_{0}=(\mu\lambda_0)^{1/n}\exp[-Y_{0}]$ 
and $u_{r+1}=(\mu\lambda_{r+1})^{1/n}\exp[-Y_{r+1}]$)
\eqn\twspotl{
  \widetilde W = u_0^n + u_{r+1}^n 
		+ \sum_{\alpha=1}^r
                \lambda_\alpha' u_0^{p_\alpha}u_{r+1}^{q_\alpha}\ ,
}
where 
\eqn\newlambds{
  \lambda_{\alpha}' = \lambda_{\alpha} 
	\;\Lambda^{1-\Delta_{\alpha}}
	\;e^{t'_{\alpha,{\rm bare}}} 
	= \lambda_{\alpha}
		\;\mu^{1-\Delta_{\alpha}}
		\;e^{t'_{\alpha,\eff}(\mu)}
\ .
}
%The massless noncompact region of the sigma model at
%large $|X|$ is thus (by the second of equations \horivafa)
%at large $Re(Y)$ and thus small $u$.
%
Note that the $\alpha^{\rm th}$ monomial in the sum
in \twspotl\  is just 
\eqn\sigpdef{
  \Sigma'_\alpha :=(\lambda_{\alpha}')^{-1}C_{\alpha\beta}\Sigma_\beta 
	= u_0^{p_\alpha}u_{r+1}^{q_\alpha} 
}
(we can also extend this to define $\Sigma'_0 = u_0^n$,
$\Sigma'_{r+1} = u_{r+1}^n$).
The scaling dimensions of these operators thus
identifies them as $\Sigma'_\alpha \propto \CT_{q_\alpha}$.

One should be careful in the use of the
`mirror transformation' \refs{\MorrisonYH,\HoriKT}.
The mirror transformation amounts to T-duality on the phase
of the $X_j$; however, in the geometrical
phase, the minimal volume cycles $\CC_\alpha$
are at $X_\alpha=0$ where this T-duality is ill-defined.
It is therefore unclear to what extent the effective
superpotential \twspotl\ will accurately capture 
the properties of the `geometrical' supersymmetric
ground states associated to the homology
of the resolved Hirzebruch-Jung space of the Higgs branch.  
We do however expect it to describe correctly 
the vacua of the Coulomb branch
which are supported away from the origin,
and it is only for this purpose that we will employ it.
One could carry out the whole analysis of massive
vacua in terms of the $\Sigma'_\alpha$
without introducing the auxiliary fields $Y_i$;
it is merely for convenience that we introduce them.

Note that the `mirror' $\IZ_n$ transformation
\eqn\mirzn{
  (u_0,u_{r+1})\sim(\omega u_0,\omega^{-p}u_{r+1})
}
leaves the effective superpotential \twspotl\ invariant --
it fixes all the $\Sigma'_\alpha$.  Indeed it is a 
{\it gauge symmetry} remnant of the duality transformation 
and therefore we should quotient the LG model by its action.  
%
%and therefore is
%a symmetry of the mirror transformation under which
%the mirror description should be identified.
%
Thus massive vacua of \twspotl\ come in orbits of length $n$ in $u$-space.
In general complex dimension $d$, the dualized theory must
be orbifolded by $(\IZ_{n})^{d-1}$.%
\foot{The $U(1)^d$ $R$-charges of the twist operators are of the form
$\coeff1n\vv_j=j\ww$
(mod $\IZ$ in each component), for $j=1,...,n-1$.
There are $d-1$ independent vectors orthogonal to $\ww$, whose rational
components are in $\coeff1n \IZ$ and define a $\IZ_n^{d-1}$
action shifting the $Y$'s, which fixes the $\Sigma_\alpha$.
This transformation therefore represents a redundancy of the 
description under which the $Y$'s should be identified.
Equation \mirzn\ represents the special case $d=2$.}

\subsec{Counting the vacua}

The effect of turning on the FI couplings $\lambda_\alpha$
is to move a subset of the critical points of \twspotl\
out to large $|u|$ for large $|\lambda|$.  
A simple way to see this is to rescale
\eqn\rescalvac{\eqalign{
  u_0 &\to zu_0 \cr
  u_{r+1} &\to zu_{r+1} \cr
  \lambda'_\alpha &\to z^{n-p_\alpha-q_\alpha}\lambda'_\alpha \cr
}}
which homogeneously rescales the twisted superpotential \twspotl;
thus, extrema of the potential scale to large $u_0$, $u_{r+1}$
at large $\lambda'$.
On the other hand, the `geometrical region' 
of the configuration space -- the Higgs branch --
remains at $|u|\sim0$ as the FI parameters are made large.
One can understand this latter property
in terms of the original variables $\Sigma_\alpha$, $X_i$;
equation \convexreg\ says that the geometrical region 
(the Hirzebruch-Jung space obtained by solving the D-term equations)
is far from the origin in the variables $X_0$, $X_{r+1}$.
The component potential term $Q_{\alpha i}^2 |\sigma_\alpha|^2|X_i|^2$
in \Xpotl\
then forces $\sigma_\alpha\sim 0$ 
and thus $u_0,u_{r+1}\sim 0$
by \sigpdef.  On the other hand, vacua with $u_0$, $u_{r+1}$
large have $|X|\sim 0$.  Thus for large $\zeta$ the vacua 
of the Higgs branch   and the vacua of the Coulomb 
branch are well separated, with a 
potential barrier between them.%
\foot{See the last paragraph of section 2 for a more 
accurate description.}

This flow to infinite separation
is the standard mechanism by which RR ground states
decouple in $\CN=2$ supersymmetric field theories;
here the decoupling has an interpretation of
decoupling certain K-theory charges from the
geometrical spacetime as one perturbs away from the 
orbifold geometry.

Let us now count how many vacua decouple to large $|u |$
in this manner.  In fact, it turns out to be 
easier to count how many remain behind (at $u_0=u_{r+1}=0$)
when all the $\lambda_\alpha$ are nonzero and generic.  
We can do this
by turning on a small additional perturbation
\eqn\deltaw{
  \delta \widetilde W= \epsilon u_0
}
in the Landau-Ginsburg potential \twspotl, 
and counting the number of independent solutions
that are near the origin (i.e. that smoothly approach $u=0$ 
as $\epsilon\to 0$) as opposed to those which are $\CO(1)$ for 
$\epsilon\to 0$.  

There are several types of scaling solutions near the
origin when we do this.  As noted at the end
of the previous section, the chiral ring generators
define a Newton boundary of monomials.  
%
%Adjacent charge
%vectorsdefine a face of a
%convex polygon known as the Newton boundary; 
%
The line passing through two adjacent 
vectors   $\vv_\alpha$, $\vv_{\alpha+1}$ 
has all monomials invariant under the mirror
$\IZ_n$ \mirzn\ lying on or above it.  
For example, in figure \newton,
the Newton boundary is the polygon with vertices
specified by the $R$-charge vectors for $\CV_{\rm\sst Y}$,
$\CT_1\equiv\Sigma'_1$, $\CT_4\equiv\Sigma'_2$, 
$\CT_7\equiv\Sigma'_3$, and $\CV_{\rm \sst X}$.

%\bigskip
%{\vbox{{\epsfxsize=2.5in
%        \nobreak
%    \centerline{\epsfbox{newton1.eps}}
%        \nobreak\bigskip
%    {\raggedright\it \vbox{
%{\bf Figure \newton.}
%{\it
%The Newton boundary of polynomials $u_0^\ell u_{r+1}^m$
%invariant under \mirzn, for $n(p)=10(3)$,
%is the solid line bounding the shaded polygon.
%The generators of the chiral ring $u_0^{p_\alpha}u_{r+1}^{q_\alpha}$,
%$\alpha=1,...,r$,
%lie on the Newton boundary.
%} }}}}
%    \bigskip}

Thus we can define a scaling $u_0\sim \epsilon^\mu$,
$u_{r+1}\sim \epsilon^\nu$ such that $\Sigma'_\alpha\sim\Sigma'_{\alpha+1}$
in their scaling, with all other $\Sigma'_\beta$ scaling as
the same or higher powers of $\epsilon$.  
This means that when analyzing
critical points of the potential that lie stably near the
origin and scale in this way, we may ignore all the other
terms in the potential and focus on the perturbing term
plus these two monomials: 
\eqn\weffi{
  \widetilde W_{\rm eff}^{\sst (\alpha)} = \epsilon u_0 +
	\lambda_\alpha' u_0^{p_\alpha}u_{r+1}^{q_\alpha} +
	\lambda_{\alpha+1}' u_0^{p_{\alpha+1}}u_{r+1}^{q_{\alpha+1}}\ .
}
Dropping all constant coefficients, the variational equations are
of the form
\eqn\vareq{
  \eqalign{
	\epsilon &= u_0^{p_\alpha-1}u_{r+1}^{q_\alpha}
			+ u_0^{p_{\alpha+1}-1}u_{r+1}^{q_{\alpha+1}} \cr
	0 &= u_0^{p_\alpha}u_{r+1}^{q_\alpha-1}
			+ u_0^{p_{\alpha+1}}u_{r+1}^{q_{\alpha+1}-1} 
}}
Note that for $\alpha=0$ the second equation does not involve
$u_{r+1}$ because $q_0=0$ (so the first term is actually absent)
and $q_1=1$.  Then the first equation is solved only 
by letting $u_{r+1}\to\infty$ (as one sees by introducing
an infinitesimal $\epsilon'$ on the LHS of the second equation)
and so we are not counting solutions that are stably near the
origin for small $\epsilon$.  Therefore we only consider
these equations for $\alpha=1,...,r$.

We solve these equations as follows: Without loss of generality
we can take $\epsilon$ real and positive; then we set
\eqn\usoln{
  u_0=e^{2\pi i\varphi_0}\,\epsilon^\mu	\quad,\qquad
  u_{r+1}=e^{2\pi i\varphi_{r+1}}\,\epsilon^\nu
}
with $\mu=\frac{q_{\alpha+1}-q_{\alpha}}{n-q_{\alpha+1}+q_{\alpha}}$,
$\nu=\frac{p_{\alpha}-p_{\alpha+1}}{n-q_{\alpha+1}+q_{\alpha}}$.
Plugging into \vareq, 
one has a solution for every $\varphi_0, \varphi_{r+1}~\mod~1$ 
such that: 
\eqn\phasecond{
  \varphi_0 (p_\alpha-1)+\varphi_{r+1} q_\alpha
	=\varphi_0(p_{\alpha+1}-1)+\varphi_{r+1} q_{\alpha+1}=0
	~{\rm mod}~1
}
One can check that for such solutions 
the implicit assumptions that $u_0$, $u_{r+1}$ are nonzero
and stably near zero are both satisfied.

Now, let us count these solutions.  We may interpret 
\phasecond\ as the defining equations for the 
lattice ${\bf L}^{\!*}$ in $\IR^2$ dual to the lattice $\bf L$ 
spanned by $\ww_1= (p_{\alpha}-1, q_{\alpha})$ and 
$\ww_2 = (p_{\alpha+1}-1, q_{\alpha+1})$.  We are only 
interested in the number of vectors in ${\bf L}^{\!*}/\IZ^2$ 
since $\varphi_0, \varphi_{r+1}$ are only defined modulo 1.  
The number of such vectors is the volume of the unit cell of 
$\bf L$, and hence  the $\alpha^{\rm th}$ scaling solution has 
$\ww_2\times\ww_1=n+q_\alpha-q_{\alpha+1}$ zeroes near the origin,
for a total of
\eqn\totalzero{
  \sum_{\alpha=1}^r (n+q_\alpha-q_{\alpha+1}) = rn+q_1-q_{r+1} = (r-1)n+1
}
critical points stably at the origin for small $\epsilon$.

But this is exactly what we were looking to find!
There are $(n-1)^2$ critical points of \twspotl\
under variation of $u_0,u_{r+1}$;
if $(r-1)n+1$ are left at the origin under the generic
perturbation by the $\lambda_\alpha$, then there
should be $(n-r-1) n$ critical points away from the 
origin even at $\epsilon=0$. These are 
arranged in $(n-r-1)$ orbits of length $n$ under
the mirror $\IZ_n$ \mirzn. Since we must   quotient by
this action we conclude there are $n-r-1$ massive vacua under
the perturbations of the superpotential
which blow up the exceptional curves of the minimal resolution.%
\foot{We have not checked explicitly that the critical points 
away from $u=0$ are nondegenerate, but expect this to be so 
generically. It does not immediately follow from nondegeneracy 
that the vacuum is massive. In order to conclude this we 
need to know about the kinetic terms for the $u$ fields. 
We are assuming that they are well approximated by 
standard kinetic terms.} 
This corresponds precisely to the number of nontrivial fractional branes
we are expecting to lose given  the K-theory of the resolved Hirzebruch-Jung space
described above. 

One might ask if the UV limit of the D-branes 
of the Coulomb branch
transform under the quantum $\IZ_n$ symmetry in representations 
that are complementary to the special representations. This is 
indeed true.  We saw in the previous section that the
additional curves of a non-minimal   resolution
of the singularity were in general associated to 
additional $\IZ_n$ representations.  We also saw that
the additional curves were blown down along RG flow
to the infrared.  What happens is that the RR ground states
associated to these representations pass onto the
Coulomb branch of the configuration space.

In terms of the mirror LG picture, even though the
term $u_0^{p_*}u_{r+1}^{q_*}$ in the potential is
above the convex hull (the Newton boundary)
set by the monomials $u_0^{p_\alpha}u_{r+1}^{q_\alpha}$
of the minimal resolution,
by suitably making the coupling $\lambda'_*$ large enough
the former term will be just as important as the latter terms.
This is the reflection in the LG picture
of the fine tuning of the FI parameters that adds an extra
segment to the boundary of the region $\CD$, as in figure \rhoflow a.
Then, in the counting of vacua of the Coulomb branch,
one should split up the perturbations controlled
by \weffi\ for $\alpha=k$ into two parts,
one for the pair $u_0^{p_k}u_{r+1}^{q_k}$, $u_0^{p_*}u_{r+1}^{q_*}$
and another for the pair
$u_0^{p_*}u_{r+1}^{q_*}$, $u_0^{p_{k+1}}u_{r+1}^{q_{k+1}}$.
The sum \totalzero\ will have an extra term, and so one
will find one more vacuum (more precisely, an orbit of \mirzn\
of length $n$ of vacua) near the origin.
 
In other words, by suitable fine tuning
the RR vacuum corresponding to the extra $-1$ curve
remains in the region of small $|u|$ where it has a
geometrical interpretation on the Higgs branch
(and the LG picture of the Coulomb branch
is not actually reliable); the associated RR
gauge field doesn't decouple from spacetime dynamics, 
at least for some range of RG scales 
where the extra curve has a size much larger than
the string scale.  Eventually, the curve shrinks away
along the RG flow and joins the other massive
LG vacua on their exodus from the Higgs branch
of the configuration space, since the
critical points of the LG potential \twspotl\
are generically controlled by the most rapidly
growing terms -- namely those on the Newton boundary.

%%%%%%%%%%%%%%%%%%%%%%%%%%%%%%%%%%%%%%%%%%%%%%%%%%%%%%%%%%%%%%%%%%%

\subsec{The `optimal' resolution}

It would then appear that one can treat any of the vacua,
massive or massless, associated to any particular perturbation
of the chiral ring of the orbifold CFT, 
and follow how it joins or leaves the 
Higgs branch of the GLSM configuration space.
The ideal starting point would employ
a $U(1)^{n-1}$ GLSM rather than $U(1)^r$, and let
fields decouple as one prescribes.  One now also has the
degrees of freedom to express the gauge
field for {\it any} nontrivial line bundle $\CR_j$
present at the orbifold fixed point, $j=1,...,n-1$,
in terms of GLSM fields via \HJconnform.  
The logical starting point uses
the D-term equations written in the diagonalized
basis \newdees, where one has a direct relation to 
scaling operators $\CT_j$ and the associated
$\IZ_n$ representations $\rho_j$. 
Naively, it would appear that one can fine tune 
so that any representation appears 
in the Higgs branch of the configuration
space by blowing up non-minimally.

An exception to this prescription arises when the R-charge vector of
the corresponding orbifold twist operator is not primitive, \ie\
is a power of another twist operator;
then the the non-primitive twist operator does not correspond
to a distinct $\IP^1$ in the resolution.
Nevertheless, these may also be put into a canonical form.
To see this, start with the D-terms put in the diagonalized form \newdees.
Whenever $(p_\alpha,q_\alpha)=k(p_\beta,q_\beta)$ for some $\beta$,
we can rewrite the $\alpha^{\rm th}$ constraint
by taking a linear combination with the $\beta^{\rm th}$ constraint as
\eqn\Dmult{
  k|X_\beta|^2-|X_\alpha|^2=-k\zeta'_\beta+\zeta'_\alpha
	\equiv\hat\zeta'_\alpha\ .
}
When $\hat\zeta'_\alpha<0$, $|X_\alpha|$ is forced to be nonzero,
and we can gauge fix the $U(1)$ action associated
to the above D-term constraint via fixing the phase of $X_\alpha$.
There is no residual gauge action and the remainder of
the theory is unaffected.  If on the other hand $\hat \zeta'_\alpha>0$,
we have $|X_\beta|$ forced to be nonzero, and the $U(1)$
action may be gauge fixed by fixing the phase of $X_\beta$.
As in section 3, this
leaves a residual $\IZ_k$ symmetry which acts on the
normal bundle to the $\beta^{\rm th}$ $\IP_1$ in the resolution chain;
it does {\it not} yield an independent $\IP^1$ of the resolved space.
A special case of this is the $\IZ_{n(1)}$ orbifold, where
blowing up to infinity via the chiral operator $W^k$
(where $W$ generates the chiral ring) leads to the
daughter space $\IC\times(\IC/\IZ_k)$ 
\MartinecTZ.

The general setup is thus indeed to start with a $U(1)^{n-1}$ GLSM,
in the diagonal basis where the $U(1)$ charges are related to the 
R-charges of the twist fields, so that the D-term constraints are
\eqn\diagmax{
  R_{\kappa i}|X_i|^2\equiv
	\frac{p_\kappa}n|Y_0|^2+\frac{q_\kappa}n|Y_n|^2-|Y_\kappa|^2
	=\zeta'_\kappa\ ,
}
with the R-charge vectors given by the full set of 
orbifold R-charges \Rchge.
There is a unique `optimal' resolution containing all primitive
vectors in the set of R-charge vectors of the orbifold twist fields
with a generalized Dynkin diagram of rank $s$.
Divide the index set $\{1,...,n-1\}$ into the subset
$\{\alpha_1,...,\alpha_s\}$ associated to this optimal resolution,
and the complement $\{\xi_1,...,\xi_{n-1-s}\}$.
Denote by $\hat B_{\kappa\mu}$ the embedding of its
generalized Cartan matrix acting nontrivially only
on the subset $\{\alpha_i\}$ within the full index space;
and denote by $\hat M_{\kappa\mu}$ the lower triangular
transformation which has the effect
of turning \diagmax\ into \Dmult\ for all the $\{\xi_j\}$
that correspond to non-primitive $R$-charge vectors.
A canonical charge matrix for the $U(1)^{n-1}$ GLSM
is thus $\hat Q=(\hat B+\hat M)\hat R$. 
We may define the geometrical FI parameters 
controlling the sizes of cycles as
$\hat\zeta_\kappa=\hat Q_{\kappa\mu}\hat\zeta'_\mu$ as usual.  
By construction, the $R$-charges of the twisted perturbations
$(q_j,p_j)$, $j=1,...,n-1$, will coincide with 
those of the orbifold \Rchge.

Because we now have a $U(1)$ gauge field corresponding to each
representation of $\IZ_{n(p)}$, one can follow
what happens to each representation as we move around the parameter space
of the twist fields.  The optimal resolution just defined keeps
them all around if we tune the $\zeta'_\kappa$ to be large and
positive in the appropriate range -- all representations
of $\IZ_{n(p)}$ can be found as canonical line bundles on
the resolved space.
RG flow results in a subset of the curves 
of the optimal resolution being blown down.
Various curves leave the resolution chain, 
and the line bundles become associated to D-branes of the 
massive vacua in the Coulomb branch 
that decouple.

Let us illustrate this procedure via the example of
$n(p)=10(3)$, for which the chiral ring $R$-charges
are depicted in figure \ring.
The minimal resolution is associated to the continued fraction 
$10/3=[4,2,2]$ with corresponding generators 
$\{\CT_1,\CT_4,\CT_7\}$.
The optimal resolution involving resolution vectors
associated to twist fields is given by the
continued fraction $10/3=[6,1,3,1,4,2]$,
with corresponding chiral ring elements (in sequence) 
$\{\CT_1,\CT_6,\CT_5,\CT_9,\CT_4,\CT_7\}$.%
\foot{Note that the generators/R-charge vectors in the resolution
chain are in order of decreasing $p_\alpha/q_\alpha$.}
Left out are the generators $\CT_2=(\CT_1)^2$,
$\CT_3=(\CT_1)^3$, and $\CT_8=(\CT_4)^2$.
The combination of the generalized Cartan matrix 
%$\hat C$ 
embedded as $\hat B$ in the $U(1)^9$, 
as well as the lower triangular matrix $\hat M$
acting on the twists that are non-primitive,
yields the charge matrix:
\eqn\maxmat{
  \hat Q=(\hat B+\hat M)\cdot\hat R =\pmatrix{
~1& -6& ~0& ~0& ~0& ~0& ~1& ~0& ~0& ~0& ~0\cr 
~0& ~2& -1& ~0& ~0& ~0& ~0& ~0& ~0& ~0& ~0\cr 
~0& ~3& ~0& -1& ~0& ~0& ~0& ~0& ~0& ~0& ~0\cr 
~0& ~0& ~0& ~0& -4& ~0& ~0& ~1& ~0& ~1& ~0\cr 
~0& ~0& ~0& ~0& ~0& -3& ~1& ~0& ~0& ~1& ~0\cr 
~0& ~1& ~0& ~0& ~0& ~1& -1& ~0& ~0& ~0& ~0\cr 
~0& ~0& ~0& ~0& ~1& ~0& ~0& -2& ~0& ~0& ~1\cr 
~0& ~0& ~0& ~0& ~2& ~0& ~0& ~0& -1& ~0& ~0\cr 
~0& ~0& ~0& ~0& ~1& ~1& ~0& ~0& ~0& -1& ~0}
%\cr
%\cr
%}}
}
Here $\hat R_{\alpha i}$ is the diagonalized matrix
from \diagmax.
Thus we see that rows 1,6,5,9,4,7 indeed form the 
standard charge matrix of the blowup.  The additional
structure from rows 2,3,8 will have the same consequence
as in the discussion after equation \Dmult;
in other words, no effect in the regime
of large negative FI parameter, while for large positive
FI parameter they will induce some extra orbifolding
of the normal bundle to a blown up curve.

%%%%%%%%%%%%%%%%%%%%%%%%%%%%%%%%%%%%%%%%%%%%%%%%%%%%%%%%%%%%%%%%%%%
%%%%%%%%%%%%%%%%%%%%%%%%%%%%%%%%%%%%%%%%%%%%%%%%%%%%%%%%%%%%%%%%%%%

\subsec{Deformed chiral ring relations}

As an aside, we note that
another use of the dual Landau-Ginsburg potential \twspotl\ 
is to derive the deformation of the chiral ring that
occurs when one perturbs away from the orbifold point
$\lambda_\alpha=0$ in the parameter space.
Note that, due to equation \area, 
the equations for a critical point can be written
\foot{We are assuming that the $u's$ are not near zero 
so that we can freely multiply these relations by $\Sigma'_{\alpha}$. } 
\eqn\crit{
  n \frac{(\Sigma'_\alpha)^{q_{\alpha+1}}}{(\Sigma'_{\alpha+1})^{q_\alpha}}
	+\sum_\beta p_\beta \Sigma'_\beta = 0
}
for all $\alpha=1,...,r$.  Therefore
\eqn\crossratio{
  \frac{(\Sigma'_\alpha)^{q_{\alpha+1}}}%
	{(\Sigma'_{\alpha+1})^{q_\alpha}}
  = \frac{(\Sigma'_{\alpha-1})^{q_\alpha}}%
	{(\Sigma'_\alpha)^{q_{\alpha-1}}}\ .
}
The consequence of this relation is (again due to equation \area,
and the fact that $\vv_\alpha=(q_\alpha,p_\alpha)$ obey the resolution
vector relations $a_\alpha\vv_\alpha=\vv_{\alpha+1}+\vv_{\alpha-1}$)
\eqn\ringrel{
  (\Sigma'_\alpha)^{a_\alpha} = 
	(\Sigma'_{\alpha-1})
		(\Sigma'_{\alpha+1})
	\quad,\qquad \alpha=2,...,r-1 .
}
This relation only holds for $\alpha=1,r$ if we use a
substitution of $\Sigma'_0$ or $\Sigma'_{r+1}$, and then
we would need a ring relation for them.  Since 
$\Sigma_0', \Sigma_{r+1}'$ correspond to nonnormalizable 
modes one should use  the relations on the $\vv_\alpha$ 
(and the fact that $q_1=1$ and $p_r=1$), 
to rewrite the equations of motion for $u_0, u_{r+1}$ as
\eqn\firstlast{
\eqalign{
  0 &= n(\Sigma'_1)^{a_1} + \sum_{\alpha=1}^r 
	p_\alpha\lambda_\alpha' \Sigma'_\alpha\Sigma'_2\cr
  0 &= n(\Sigma'_r)^{a_r} + \sum_{\alpha=1}^r 
	q_\alpha\lambda_\alpha' \Sigma'_\alpha\Sigma'_{r-1}\ .
}}
This pair of equations, together with \ringrel,
are some (but not all) of the deformed chiral ring relations.

%%%%%%%%%%%%%%%%%%%%%%%%%%%%%%%%%%%%%%%%%%%%%%%%%%%%%%%%%%%%%%%
%%%%%%%%%%%%%%%%%%%%%%%%%%%%%%%%%%%%%%%%%%%%%%%%%%%%%%%%%%%%%%%

\newsec{The spacetime effective action}

The decoupling from the Higgs branch
of D-branes and RR fields witnessed above
should be reflected in the structure of the spacetime
effective action.  We will propose such an effective
action in this section.  But first we will discuss
the effect of the GSO projection on the spectrum and
dynamics of the $\IC^2/\IZ_{n(p)}$ orbifolds.

%%%%%%%%%%%%%%%%%%%%%%%%%%%%%%%%%%%%%%%%%%%%%%%%%%%%%%%%%%%%%%%

\subsec{GSO projections and RR fields}

The chiral ring of the orbifold contains the set of BPS protected
twist operators \znptwist, which are holomorphic under the
natural choice of complex structure for the $\IC^2$ coordinates
$X$, $Y$.  However, there is another ring
\eqn\caring{
  \Sigma_{j/n}^{\sst(X)}(\Sigma_{1-\{jp/n\}}^{\sst(Y)})^* 
}
which is BPS under a different linear combination
$G_\X+G^*_\Y$ of the supersymmetry currents of the component
theories, the one natural to the opposite complex structure
obtained via $Y\to Y^*$.
We can call the ring of these latter operators
the $(c_\X,a_\Y)$ ring, and the ring of operators \znptwist\
the $(c_\X,c_\Y)$ ring.

The type 0 theory that we have been discussing
in fact contains both the $(c_\X,c_\Y)$ and $(c_\X,a_\Y)$ rings.
The type II GSO projection demands invariance under 
\eqn\gsoproj{
H_1\to H_1+p\pi\quad,\qquad H_2\to H_2-\pi
}
where $H_i$ are the bosonized worldsheet fermions; 
this keeps some of each ring, namely
the $(c_\X,c_\Y)$ states with $[jp/n]\in 2\IZ+1$
and the $(c_\X,a_\Y)$ states with $[jp/n]\in 2\IZ$
(here $[\xi]=\xi-\{\xi\}$ denotes the integer part of $\xi$).
It is only for the supersymmetric orbifold that the
entire $(c_\X,a_\Y)$ ring is projected out and
the entire $(c_\X,c_\Y)$ ring is preserved by the GSO projection.

The fact that some of the generators of the
$(c_\X,c_\Y)$ ring are projected out means that
one is obstructed in the type II theory from fully
resolving the singularity using K\"ahler deformations
alone.  It is for this reason that we have focussed
our attention on the type 0 theory.%
\foot{A. Adams has suggested to us that
one might be able to resolve the singularity fully
in the type II theory
by employing the $(c_\X,a_\Y)$ operators, which deform
the algebraic equations embedding the singularity in $\IC^{\ell+2}$
(\cf\ \MartinecTZ).  However, since these operators are
BPS under a different choice of complex structure,
they are not protected from renormalization within
the same scheme that protects the $(c_\X,c_\Y)$ operators.
Thus it is difficult to follow their effect under 
the finite deformation required to obtain a geometrical 
picture of the target space.}

Spectral flow in the $\CN=2$ U(1) R-charge generates
a RR ground state, and associated RR gauge field,
for every allowed element of each ring.  
Thus in the type II theory, there are
only $n-1$ RR bispinor fields, 
some associated to the $(c_\X,c_\Y)$ ring
and some to the $(c_\X,a_\Y)$ ring.
In the type 0 theory there are twice as many RR gauge fields
coming from the twisted sectors as in type II, 
because one keeps both parities of spinor in forming
the RR bispinors.

%%%%%%%%%%%%%%%%%%%%%%%%%%%%%%%%%%%%%%%%%%%%%%%%%%%%%%%%%%%%%%%

\subsec{The spacetime effective action}

We now come to the question of the effective action
for the RR gauge fields coupling to the disappearing
fractional branes.  The disappearance at the IR fixed point
of the topological charge that they couple to 
suggests that these fields decouple from the effective
spacetime dynamics.  There are two canonical mechanisms
for this decoupling: 
Spontaneously symmetry breaking by the Higgs mechanism,
giving the gauge fields a large mass; or some sort
of confinement mechanism (sometimes called `classical confinement'
due to the fact that we are in tree level string theory).

The Higgs mechanism would require a condensation
of D-branes, which are the only objects charged under
the RR-gauge symmmetry.  However, in the massive LG vacua
that we have seen are associated to the decoupling RR ground
states, D-branes only get heavier as we move away from
the UV orbifold fixed point.  This Higgs mechanism is ruled out.

The fact that the decoupling generators of the K-theory
lattice are associated to massive vacua of the GLSM
is rather reminiscent of a similar situation in open string RG flows
\HarveyNA, where the classical confinement mechanism occurs.
Consider the open string tachyon in the $Dp$-$\bar Dp$ system.
In the wordsheet RG approach to tachyon condensation,
the tachyon condensate to a $D(p-2)$ brane appears
as a boundary mass term.  Under the flow, all boundary operators
are flowing to the identity operator in the infrared;
there are no physical excitations of the brane system
away from the massive minimum of the effective potential.
One is thus led to conjecture \SenXM\
a form of the effective action for the open string degrees of freedom
\eqn\seneffact{
  \CS^{\rm open}_\eff=\int d^{p+1}\!x\;f(T)[F_{\rm open}^2+\ldots]\ ,
}
where $f(T)\to 0$ as the tachyon condenses.  This proposal
has been verified in simple examples 
\refs{\GerasimovZP,\KutasovQP}.

The picture of the RR gauge fields which decouple under
the closed string tachyon perturbations considered here
is quite similar.  The RR gauge fields associated to the massive
vacua which are decoupling are such that all their excitations
are flowing to the identity operator in the IR of the flow.
The RG flows of the GLSM suggest a conjecture analogous to \seneffact\
for the RR gauge fields coupling to the disappearing
K-theory charges:
\eqn\RRdecouple{
  \CS^{\sst RR}_\eff=\int d^6 \! x \sum_{i=1}^{n-1} 
	f_i(T)[(F_{RR}^{(i)})^2+\ldots]\ ,
}
In perturbation theory $f_i(T) \sim 1 + \CO(T)$
(for more precise formulae, see \HarveyWM). 
However, nonperturbatively, it must be that   
$f_i(T)\to 0$ as the tachyon condenses
for those gauge fields whose charges disappear.%
\foot{The effective action of this closed string theory is 
actually an integral over 10 dimensions. However, the couplings to 
the RR fields associated with the twisted sectors are weighted by 
wavefunctions which fall off roughly as
$\sim e^{-r/\ell_{string}} $ in the 
directions transversal to the orbifold. Thus, at low energies and 
long distances, the effective action is an integral over the orbifold 
fixed point locus. }

The mechanism of decoupling of gauge charge, and excitations
that couple to it on localized unstable objects, 
would thus appear to be rather universal in perturbative string theory.

%%%%%%%%%%%%%%%%%%%%%%%%%%%%%%%%%%%%%%%%%%%%%%%%%%%%%%%%%%%%%%%%%%%
%%%%%%%%%%%%%%%%%%%%%%%%%%%%%%%%%%%%%%%%%%%%%%%%%%%%%%%%%%%%%%%%%%%
%%%%%%%%%%%%%%%%%%%%%%%%%%%%%%%%%%%%%%%%%%%%%%%%%%%%%%%%%%%%%%%
%%%%%%%%%%%%%%%%%%%%%%%%%%%%%%%%%%%%%%%%%%%%%%%%%%%%%%%%%%%%%%%

%%%%%%%%%%%%%%%%%%%%%%%%%%%%%%%%%%%%%%%%%%%%%%%%%%%%%%%%%%%%%%%%%%%

\newsec{Discussion}

What general lessons can we draw from the above considerations? 
Perhaps the most important one is that $K$-theory might continue 
to play an interesting role in {\it closed string} tachyon 
decay. One general viewpoint on the relation of K-theory and 
D-branes is that K-theory is an invariant of {\it boundary} 
RG flow, for a fixed bulk CFT. In general, there
is no particular reason to think that $K$-theory of spacetime
should be an invariant of bulk RG flow.  However, $\CN=2$
worldsheet supersymmetry gives additional structure
that allows one to relate D-branes and K-theory
in a way that is preserved under bulk RG flow.
Furthermore, the $\CN=2$ preserving localized tachyon perturbations 
lead to a controlled family of closed string RG flows. 
In this case one might expect to be able to determine 
the ``fate'' of the K-theory charges. 
This is what we have accomplished in the present paper.  

In essence, the topology of the target space 
is defined by specifying precisely
the UV fixed point theory on the worldsheet that describes
the unstable, tachyonic vacuum of the closed string theory.
One might be able to regard this as a manifestation of
UV/IR duality (the global structure of spacetime is related
to short-distance structure on the worldsheet).
For example, one could have considered the Hirzebruch-Jung space
with an extra million non-minimal blowups to be the UV
theory one starts with; this will have a K-theory lattice
whose rank is increased by a million over that of the minimal
resolution, then at some crossover scale the extra
curves will blow down and the extra structure decouples,
leaving behind the minimal resolution and its K-theory
lattice of rank $r+1$.  Or one can consider the orbifold
fixed point which also flows to the same IR theory,
which will in general have a K-theory lattice smaller than
the above in rank, but still larger than that of the IR 
Hirzebruch-Jung space.  Different UV theories having different
K-theory can flow to the same Higgs branch geometry
in the IR.  The full UV theory keeps track of
all the topology, some of which moves to 
non-geometrical branches of the configuration space
along the flow to the IR. 
The Higgs branch of the IR limit contains only
the topology of the $r$ curves of the minimal resolution.

It is interesting to contrast our discussion of D-branes in 
the massive vacua with the paper of Hori, Iqbal, and Vafa \HoriCK. 
These authors studied D-branes on compact toric manifolds with 
$c_1(X)>0$. 
Such sigma models are good UV fixed points for the massive 
D-branes. Under RG flow to the IR these manifolds shrink to zero size, 
and the IR description of the theory is a Landau-Ginzburg 
orbifold. The latter is more appropriately described by the 
Coulomb branch vacua. Thus, one should speak of {\it either} the Higgs branch, 
{\it or} the Coulomb branch, and the mirror correspondence 
of D-branes discussed in \HoriCK\ is a correspondence between 
the IR and UV description of the ``same'' branes. In the 
examples studied in this paper the 
Hirzebruch-Jung manifold has $c_1(X)<0$, 
and appears in the IR, not the UV region of the theory. 
Hence, in the flow to the IR one is forced to 
discuss {\it both} the Higgs and the Coulomb branches.

The phenomenon of open string tachyon condensation has been 
the focus of some very interesting investigations in string 
field theory in the past few years.  A corresponding theory 
for closed string tachyon condensation is glaringly absent.  
The spacetime picture we have advocated for the disappearance
of RR U(1) gauge fields
under localized closed string tachyon condensation
suggests a natural set of conjectures to which one might try to
apply the techniques of closed string field theory. 

%%%%%%%%%%%%%%%%%%%%%%%%%%%%%%%%%%%%%%%%%%%%%%%%%%%%%%%%%%%%%%%%%%%

\bigskip  
\noindent{\bf Acknowledgements:}  
We would like to thank  J. Harvey and D. Kutasov 
for continued discussions on localized closed string 
tachyon condensation. (We are especially grateful to DK 
for his forthright comments on an early draft of this paper.) 
We would also like to thank 
B. Acharya, A. Adams, E. Diaconescu, 
M. Douglas, B. Florea, R. Gregory, H. Liu, D. Morrison, R. Plesser,   
M. Reid, and W. Taylor  for useful discussions.    
The support and hospitality of the Aspen Center for Physics  
during the initial stages of this work is gratefully appreciated.
The work of E.M. is supported in part by DOE grant DE-FG02-90ER40560. 
The work of G.M. is supported in part by DOE grant DE-FG02-96ER40949.

\appendix{A}{Some properties of the matrix $\CN$}

\subsec{Proof of \niceident} 

We wish to show that
\eqn\appone{ 
  \CN^{-1}_{\alpha\gamma} Q_{\gamma,\beta+1}
        = |X_{\beta+1}|^{-2} \delta_{\alpha\beta} 
}
on the surface $X_\beta = X_{\beta-1} = 0$.
To do this, first note that $\CN_{\alpha\gamma}$ is block diagonal
on this surface, namely $\CN_{\alpha\gamma}=0$
for $\alpha<\beta$, $\gamma\ge\beta$.
This implies that  $\CN^{-1}_{\alpha\gamma} Q_{\gamma,\beta+1} = 0$
for $\alpha<\beta$.  We can thus concentrate on the
lower right block of entries for $\alpha,\gamma\ge\beta$.
Call this block $\tilde \CN_{\alpha\gamma}$,
and the corresponding block of $Q$ we will call
$\tilde Q_{\alpha i}$ for $\alpha,i\ge\beta$.

It is not hard to show that
\eqn\apptwo{
   \tilde \CN_{\alpha\gamma} = L \cdot D \cdot U 
}
where
\eqn\appthree{
  L_{\alpha\gamma}=\tilde Q_{\alpha,\gamma+1}          \qquad
   U_{\alpha\gamma}=\tilde Q_{\alpha+1,\gamma}          \qquad
   D_{\alpha\gamma}=diag(|X_{\beta+1}|^2,...,|X_{r+1}|^2) 
}
are lower triangular, upper triangular, and diagonal matrices,
respectively.  Furthermore,
\eqn\appfour{
  (L^{-1} \cdot Q)_{\alpha,\gamma} = \delta_{\alpha,\gamma-1} 
}
for $\alpha\ge \beta$, $\gamma\ge\beta+1$.
Then we have
\eqn\appfive{
  (\CN^{-1} \cdot Q) = U^{-1} \cdot D^{-1} \cdot L^{-1} \cdot Q 
}
and the RHS is manifestly upper triangular in the relevant block.
This means that
\eqn\appsix{
  (\CN^{-1} \cdot Q)_{\beta,\beta+1} = |X_{\beta+1}|^{-2}        \qquad
   (\CN^{-1} \cdot Q)_{\alpha,\beta+1} = 0 \quad,\qquad \alpha>\beta 
}
which is what we were to show.

\subsec{An explicit inverse for $\CN$} 

It is possible to give an explicit inverse for the matrix $\CN$. 
While it is not used in the text, this formula is slightly nontrivial, 
and might prove useful in future investigations. So we give it here. 

In order to invert $\CN$ consider the matrix 
 $T= D^{-1} C^{-1} \CN C^{-1} D^{-1}$ where
\eqn\appseven{
  D_{\alpha\beta} = \delta_{\alpha\beta} \vert X_{\alpha}\vert\quad .
}
This matrix is of the form 
\eqn\gentee{
T = 1 + v_1 v_1^T + v_2 v_2^T 
}
where 
\eqn\defveew{
\eqalign{
(v_1)_{\alpha} & = 
{\vert X_0\vert \over \vert X_{\alpha} \vert} {p_\alpha\over n} , \cr
(v_2)_{\alpha} & = 
{\vert X_{r+1}\vert \over \vert X_{\alpha} \vert} {q_\alpha\over n} 
\quad .
}}
The inverse of a matrix of the form \gentee\ is 
\eqn\genteeii{
\eqalign{
& T^{-1} = 1 - {1\over \Delta}\biggl(
( 1 + v_2^2) v_1 v_1^T 
 -  (1+v_2^2)  v_2 v_2^T   +   
(v_1\cdot v_2)   (v_1 v_2^T + v_2 v_1^T)\biggr) \cr}
}
where it is convenient to introduce 
$\Delta:= 1+ v_1^2 + v_2^2 + v_1^2 v_2^2 - (v_1\cdot v_2)^2$.
Applying \genteeii\  to our case we find: 
\eqn\invertenn{
\eqalign{
\CN^{-1}_{\alpha\beta} &= 
\sum_{\gamma} {C^{-1}_{\alpha\gamma} 
C^{-1}_{\beta\gamma} \over \vert X_{\gamma}\vert^2} 
- {1\over n^2 \Delta} \Biggl[ (1+v_2^2) 
\vert X_0\vert^2 U_{\alpha} U_{\beta} + 
(1+v_1^2) \vert X_{r+1}\vert^2 V_{\alpha} V_{\beta}  \cr
&\qquad\qquad\qquad\qquad\qquad\qquad
- {\vert X_0\vert^2 \vert X_{r+1}\vert^2\over n^2} 
(\sum_\gamma {p_\gamma q_{\gamma}\over 
\vert X_{\gamma}\vert^2} ) \left(U_\alpha V_{\beta} 
+ V_{\alpha} U_{\beta}\right) \Biggr] \cr}
}
where we need to introduce vectors: 
\eqn\axudfi{
\eqalign{
U_\alpha & := \sum_{\gamma} 
{C^{-1}_{\alpha \gamma}p_\gamma\over \vert X_{\gamma}\vert^2}\cr
V_\alpha & := 
\sum_{\gamma}{C^{-1}_{\alpha\gamma}q_\gamma\over\vert X_{\gamma}\vert^2}
\quad .
}}
Using this formula it is possible to give 
a completely explicit formula for the K\"ahler quotient 
metric on $\CS_{\zeta}/U(1)^r$.

\listrefs

\bye